\documentclass[journal]{IEEEtran}
\usepackage{amsfonts}
\usepackage{enumerate}
\usepackage{epsfig}
\usepackage{amssymb}
\usepackage{pifont}

\usepackage{color}

\ifCLASSOPTIONcompsoc
  \usepackage[caption=false,font=normalsize,labelfont=sf,textfont=sf]{subfig}
\else
  \usepackage[caption=false,font=footnotesize]{subfig}
\fi

\usepackage[switch,pagewise]{lineno}
\definecolor{brightmaroon}{rgb}{0.76, 0.13, 0.28}

\usepackage[hyphens]{url}
\usepackage[hidelinks]{hyperref}
\hypersetup{breaklinks=true}
\usepackage{cite}
\usepackage{acro}
\usepackage{amsthm}
\usepackage{amsmath}
\usepackage{graphicx}
\usepackage{multirow}
\usepackage{epstopdf}
\usepackage{hyperref}

\theoremstyle{definition}

\ifCLASSINFOpdf
\else
\fi
\acsetup{list-style=tabular}

\DeclareAcronym{iot}{
  short = IoT ,
  long  = Internet of Things ,
  class = abbrev
}
\DeclareAcronym{sdr}{
  short = SDR ,
  long  = Software-defined Radio ,
  class = abbrev
}
\DeclareAcronym{asic}{
  short = ASIC ,
  long  = Application-specific Integrated Circuit,
  class = abbrev
}
\DeclareAcronym{fpga}{
  short = FPGA ,
  long  = Field Programmable Gate Array,
  class = abbrev
}
\DeclareAcronym{nfv}{
  short = NFV ,
  long  = Network Functions Virtualization ,
  class = abbrev
}
\DeclareAcronym{fft}{
  short = FFT ,
  long  = Fast Fourier Transform ,
  class = abbrev
}
\DeclareAcronym{gpp}{
  short = GPP ,
  long  = General Purpose Processor ,
  class = abbrev
}
\DeclareAcronym{dsp}{
  short = DSP ,
  long  = Digital Signal Processor ,
  class = abbrev
}
\DeclareAcronym{dac}{
  short = DAC ,
  long  = Digital-to-Analog Converter ,
  class = abbrev
}
\DeclareAcronym{adc}{
  short = ADC ,
  long  = Analog-to-Digital Converter ,
  class = abbrev
}
\DeclareAcronym{snr}{
  short = SNR ,
  long  = Signal-to-Noise Ratio ,
  class = abbrev
}
\DeclareAcronym{gpu}{
  short = GPU ,
  long  = Graphic Processing Units ,
  class = abbrev
}
\DeclareAcronym{flops}{
  short = FLOPS ,
  long  = Floating Point Operations Per Second ,
  class = abbrev
}
\DeclareAcronym{hls}{
  short = HLS ,
  long  = High-level Synthesis ,
  class = abbrev
}
\DeclareAcronym{soc}{
  short = SoC ,
  long  = System-on-Chip ,
  class = abbrev
}
\DeclareAcronym{cuda}{
  short = CUDA ,
  long  = Compute Unified Device Architecture ,
  class = abbrev
}
\DeclareAcronym{usrp}{
  short = USRP ,
  long  = Universal Software Radio Peripheral ,
  class = abbrev
}

\begin{document}
%
\title{\textsf{Software-defined Radios:\\ Architecture, State-of-the-art, and Challenges}}
%
%
%


\author{\IEEEauthorblockN{Rami Akeela, and
Behnam Dezfouli
}\\
\IEEEauthorblockA{Internet of Things Research Lab, Department of Computer Engineering, Santa Clara University, USA
\\
Email: rakeela@scu.edu, bdezfouli@scu.edu
}}

%
%

\markboth{\textsc{Internet of Things Research Lab, Department of Computer Engineering, Santa Clara University, USA }--- March~2018}%
{Akeela \MakeLowercase{\textit{et al.}}: Software-defined Radios: Architecture, State-of-the-art, and Challenges}

%



\maketitle

\begin{abstract}

Software-defined Radio (SDR) is a programmable transceiver with the capability of operating various wireless communication protocols without the need to change or update the hardware. 
Progress in the SDR field has led to the escalation of protocol development and a wide spectrum of applications, with more emphasis on programmability, flexibility, portability, and energy efficiency, in cellular, WiFi, and M2M communication. 
Consequently, SDR has earned a lot of attention and is of great significance to both academia and industry.
SDR designers intend to simplify the realization of communication protocols while enabling researchers to experiment with prototypes on deployed networks. 
This paper is a survey of the state-of-the-art SDR platforms in the context of wireless communication protocols. 
We offer an overview of SDR architecture and its basic components, then discuss the significant design trends and development tools. 
In addition, we highlight key contrasts between SDR architectures with regards to energy, computing power, and area, based on a set of metrics. 
We also review existing SDR platforms and present an analytical comparison as a guide to developers. 
Finally, we recognize a few of the related research topics and summarize potential solutions.

\end{abstract}


%
\IEEEpeerreviewmaketitle

\section{Introduction}

\IEEEPARstart{A}{dvances} in wireless technologies have altered consumers communication habits. Wireless technologies are an essential part of users’ daily lives, and its effect will become even larger in the future. In a technical report, the World Wireless Research Forum (WWRF) has predicted that 7 trillion wireless devices for 7 billion people will be deployed by 2020 \cite{wwrf09}. When these huge numbers of devices are connected to the Internet to form an Internt of Things (IoT) network, the first challenge is to adjust the basic connectivity and networking layers to handle the large numbers of end points. There is an increasing number of wireless protocols that have been developed, such as ZigBee, BLE, LTE, and new WiFi protocols for loT and Machine-to-Machine (M2M) communication purposes due to different demanding requirements, one of which is high energy efficiency \cite{Al-Fuqaha2015}. Wireless standards, in general, are adapting quickly in order to accommodate different user needs and hardware specifications \cite{IEEE,3GPP}. This has called for a transceiver design with the ability to handle several protocols, including the existing ones and those being developed. In order to accomplish this task, one needs to realize the protocols' need for a \textit{flexible}, \textit{re-configurable}, and \textit{programmable} framework.  
 
Both consumer enterprise and military frameworks have a need for programmable platforms. Programmability is of central 
significance for designers in the industry due to wireless protocols that advance rapidly and consistently, hence requiring a hardware that can keep up with the evolution. For example, the authors in \cite{Bansal2012} proposed a platform, OpenRadio, for programming both PHY and MAC layers while offering a high level of abstraction. Rather than including yet another equipment to deal with a new standard or recurrence band, the equipment of a formerly introduced platform can adjust to the particulars of another standard. In a military scenario, for example, the needs of these platforms change in light of terrible conditions that develop during a mission, which may not have been predicted when designed initially, leading to the development and utilization of new protocols. 

\ac{sdr} is a technology for radio communication. This technology is based on software-defined wireless protocols, as opposed to hardware-based solutions. This translates to supporting various features and functionalities, such as updating and upgrading through reprogramming, without the need to replace the hardware on which they are implemented. This opens the doors to the possibility of realizing multi-band and multi-functional wireless devices. 

The driving factors for the high demand of SDR include network interoperability, readiness to adapt to future updates and new protocols, and more importantly, lower hardware and development costs. 
In a report \cite{reportbuyer2016}, the SDR market is projected to be more than USD 29 billion by the year 2021. 
Global Industry Analysts \cite{sdr_report2017} highlights some of the market trends for SDR as follows: (i) increasing interest from the military sector in building communication systems and large-scale deployment in developing countries, (ii) growing demand for public safety and disaster preparedness applications, and (iii) building virtualized base stations (BSs). 
SDRs are also ideal for developing future space communications \cite{Paillassa,Angeletti2008,Angeletti2014}, Global Navigation Satellite System (GNSS) sensors \cite{Seo2011}, Vehicle-to-Vehicle (V2V) communication \cite{Xiang2015,Singh2014a,Kloc2017B}, and IoT applications \cite{Chen2016,Park2017}, where relatively small and low-power SDRs can be utilized. 

SDRs are implemented through employing various types of hardware platforms, such as General Purpose Processors (GPPs), Graphics Processing Units (GPUs), Digital Signal Processors (DSPs), and Field Programmable Gate Arrays (FPGAs). 
Each of these platforms is associated with their own set of challenges. 
Some of these challenges are: utilizing the computational power of the selected hardware platform, keeping the power consumption at a minimum, ease of design process, and cost of tools and equipment. 
The research community and industry have both developed SDRs that are based on the aforementioned hardware platforms. 
A few examples include USRP \cite{usrp}, Sora \cite{Tan2011}, Atomix \cite{Bansal2015}, Airblue \cite{Ng2010}, and WARP \cite{warp}. 
Each SDR is unique with regards to the design methodology, development tools, performance, and end application.

In this paper, we first present an overview of SDR architecture as well as the analog and digital divides of the system and interconnection of components. 
We then introduce the criteria based on which the different hardware platforms are classified. 
We thoroughly examine the architecture and design approaches employed by these hardware platforms and present their strengths and weaknesses in the context of SDR implementation. 
Furthermore, we provide an analytical comparison of hardware platforms as a guide for design decision making.  
Moreover, we discuss the use of respective development tools and present a summary to help explain their functionalities and the platforms they support. 
Afterwards, we review the SDR platforms developed by both industry and academia, and analyze and compare them using the criteria discussed before. 
Finally, we identify current challenges and the open research topics related to future SDR development.

This paper is organized as follows: Section \ref{sdr} provides a description of SDR architecture as well as the classification process used to summarize the various design approaches adopted. 
Section \ref{tools} lists some of the corresponding development tools and platforms. 
Section \ref{plat} presents an analysis and comparison of the commercially and academically developed SDR platforms. 
Research questions and future trends are highlighted in Section \ref{related}.
Section \ref{old} presents an analysis of the existing literature on SDR surveys. 
We conclude the paper in Section \ref{conc}.
A list of key abbreviations used in this paper can be found in Table \ref{acro}.

\begin{table}[]
\centering
\caption{Key Abbreviations}
\label{acro}
\begin{tabular}{ll}
\hline
\textbf{ADC}   & Analog-to-Digital Converter             \\
\textbf{ASIC}  & Application-specific Integrated Circuit \\
\textbf{BS}    & Base Station\\
\textbf{CUDA}  & Compute Unified Device Architecture     \\
\textbf{DAC}   & Digital-to-Analog Converter             \\
\textbf{DSP}   & Digital Signal Processor                \\
\textbf{FFT}   & Fast Fourier Transform                  \\
\textbf{FLOPS} & Floating Point Operations Per Second    \\
\textbf{FPGA}  & Field Programmable Gate Array           \\
\textbf{GPP}   & General Purpose Processor               \\
\textbf{GPU}   & Graphics Processing Unit                \\
\textbf{HLS}   & High Level Synthesis                    \\
\textbf{NFV}   & Network Function Virtualization       \\
\textbf{SDR}   & Software-defined Radio                  \\
\textbf{SDR}   & Software-defined Network                  \\
\textbf{SNR}   & Signal-to-noise Ratio                   \\
\textbf{SoC}   & System on Chip                          \\
\textbf{USRP}  & Universal Software Radio Peripheral     \\\hline
\end{tabular}
\end{table}


%
%
%
%

\section{Concepts and Architecture}
\label{sdr}

In this section, we examine the general architecture of SDRs, their main components, and their processing requirements. As explained in the previous section, SDRs play a vital role in wireless standard development due to their flexibility and ease of programmability. This is due to the fact that most digital signal processing, and digital front end, which includes channel selection and modulation/demodulation, take place in the digital domain. This is usually performed in software running on processors, such as GPPs and DSPs. However, it can also run on programmable hardware, i.e., FPGAs. 

%
\begin{figure}[!]
\centering
\includegraphics[width=1\linewidth]{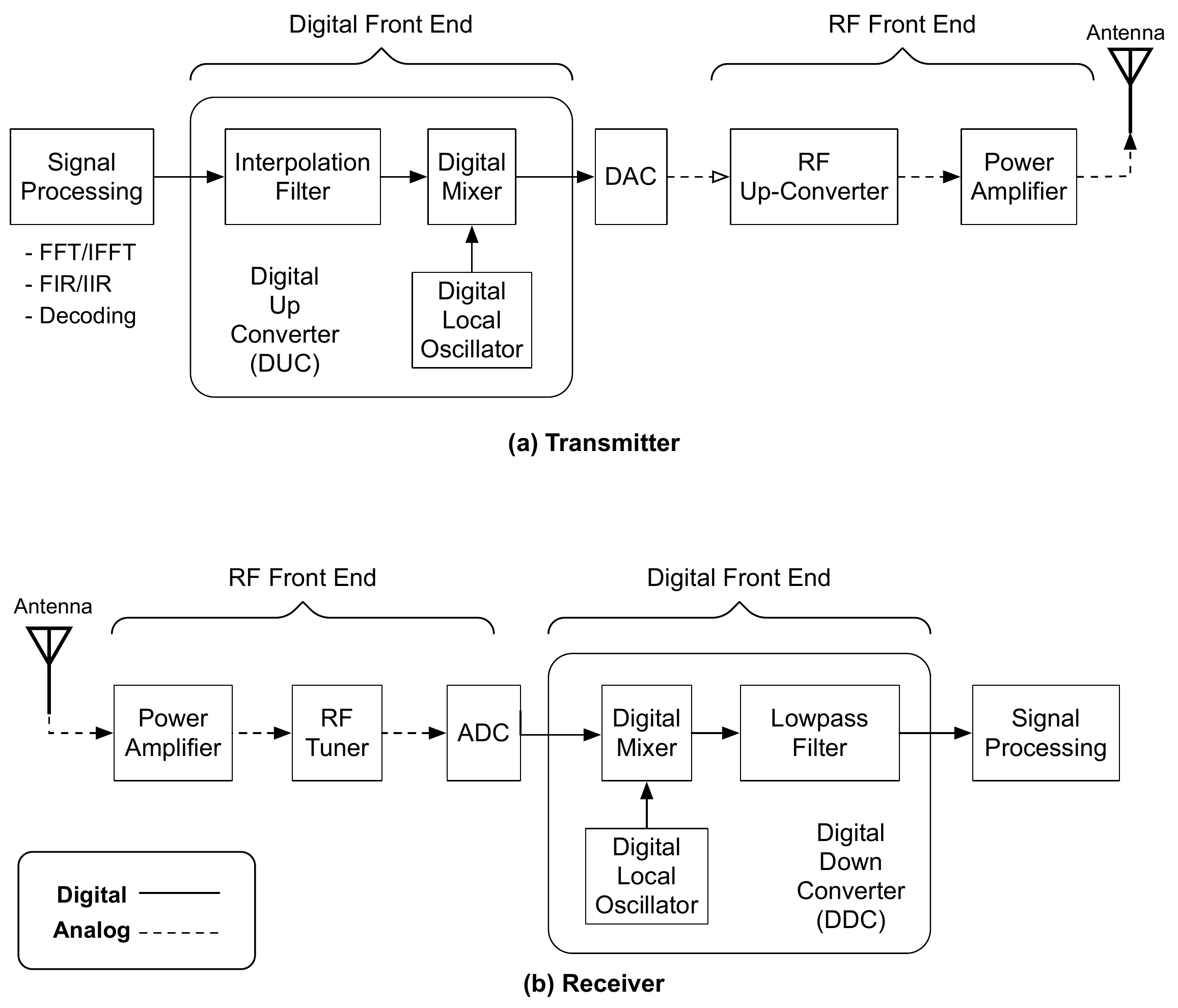}
\caption{SDR architecture. Sub-figure (a) shows SDR from a receiver's point of view, and sub-figure (b) shows SDR from a transmitter's point of view.}
\label{sdr_combined}
\end{figure}


%

In general, from the transmitter point of view, a baseband waveform needs to be produced, then an Intermediate Frequency (IF) waveform, generate an RF waveform, then send it through the antenna. 
From the receiver point of view, this RF signal is sampled and de-modulated, then decoded. 
To provide more details to the process, we study the receiving end of the system as follows.

The RF signal from the antenna is amplified, with a tuned RF stage that amplifies a range of the frequency band. This amplified RF signal is then converted to an analog IF signal. The Analog-to-Digital Converter (ADC) digitizes this IF signal into digital samples. Then, it is fed into a mixer stage. The mixer has another input coming from a local oscillator with a frequency set by the tuning control. 
The mixer then translates the input signal to a baseband. 
The next stage is essentially a FIR filter that permits only one signal. 
This filter limits the signal bandwidth and acts as a decimating low-pass filter. 
The digital down-converter includes a large number of multipliers, adders, and shift-registers, in the hardware in order to accomplish the aforementioned tasks.
Next, the signal processing stage performs tasks such as de-modulation and decoding. 
This stage is typically handled by a dedicated hardware like an Application Specific Integrated Circuit (ASIC) or other programmable alternatives like FPGA or DSP \cite{Haghighat2002a}.

\label{arch}

As shown in Figure \ref{sdr_combined} (a) and (b), at a high level, a typical SDR transceiver consists of the following components: Signal Processing, Digital Front End, Analog RF Front End, and an antenna.

\subsubsection{Antenna}
SDR platforms usually employ several antennas to cover a wide range of frequency bands \cite{rohde2017}. 
Antennas are often referred to as "intelligent" or "smart" due to their ability to select a frequency band and adapt with mobile tracking or interference cancellation \cite{Rouphael2009,Haghighat2002a}. 
In the case of SDRs, an antenna needs to meet a certain list of requirements such as self-adaptation (i.e., flexibility to tuning to several bands), self-alignment (i.e., beamforming capability), and self-healing (i.e., interference rejection) \cite{Rouphael2009}.

\subsubsection{RF Front End}
This is a RF circuitry that its main function is to transmit and receive the signal at various operating frequencies. 
Its other function is to change the signal to/from the Intermediate Frequency (IF). 
The process of operation is divided into two, depending on the direction of the signal (i.e., Tx or Rx mode):
\begin{itemize}
\renewcommand\labelitemi{--}
\item In the transmission path, digital samples are converted into an analog signal by the Digital-to-Analog Converter (DAC), which in turn feeds the RF Front End. 
This analog signal is mixed with a preset RF frequency, modulated, and then transmitted.
\item In the receiving path, the antenna captures the RF signal. 
The antenna input is connected to the RF Front End using a matching circuitry to guarantee an optimum signal power transfer. 
It then passes through a Low Noise Amplifier (LNA), which resides in a close proximity to the antenna, to amplify weak signals and minimize the noise level.
This amplified signal, with a signal from the Local Oscillator (LO), are fed into the mixer in order to down convert it to the IF \cite{Carr2001}.
\end{itemize}
\subsubsection{Analog-to-Digital and Digital-to-Analog Conversion}

The DAC, as mentioned in the previous section, is responsible for producing the analog signal to be transmitted from the digital samples. On the receiver side, the ADC resides, and is an essential
component in radio receivers. The ADC is responsible for converting continuous-time signal to a discrete-time, binary-coded signals. ADC performance can be described by various parameters \cite{Walden1999,Hentschel1999} including: (i) Signal-to-Noise Ratio (SNR): the ratio of signal power to noise power in the output, (ii) resolution: number of bits per sample, (iii) Spurious-free Dynamic Range (SFDR): the strength ratio of the carrier signal to the next strongest noise component or spur, and (iv) power dissipation.
Advances in SDR development have provided momentum for ADC performance improvements. 
For example, since ADC's power consumption affects the lifetime of battery-powered SDRs, more energy efficient ADCs have been developed \cite{Bowick2011}.
 
 
\subsubsection{Digital Front End}

The Digital Front End has two main functions \cite{Hentschel1999}:
\begin{itemize}
\renewcommand\labelitemi{--}
\item Sample Rate Conversion (SRC), which is a functionality to convert the sampling from one rate to another. This is necessary since the two communication parties must be synchronized.   
\item Channelization, which includes up/down conversion in the transmitter and receiver side, respectively. 
It also includes channel filtering, where channels that are divided by frequency are extracted. Examples include interpolation and low-pass filters, as can be observed in Figure \ref{sdr_combined}.
\end{itemize}
In a SDR transceiver, the following tasks are executed in the digital front end:
\begin{itemize}
\renewcommand\labelitemi{--}
\item In the transmitting side (Figure \ref{sdr_combined}(a)), the Digital Up Converter (DUC) translates the aforementioned baseband signal to IF. 
The DAC that is connected to the DUC then converts the digital IF samples into an analog IF signal. 
Afterwards, the RF up-converter converts the analog IF signal to RF frequencies. 
\item In the receiving side (Figure \ref{sdr_combined}(b)), the ADC converts the IF signal into digital samples. 
These samples are subsequently fed into the next block, which is the Digital Down Converter (DDC). 
The DDC includes a digital mixer and a numerically-controlled oscillator. 
DDC extracts the baseband digital signal from ADC, and after being processed by the Digital Front End, this digital baseband signal is forwarded to a high-speed digital signal processing block \cite{sadiku2004}.
\end{itemize}

\subsubsection{Signal Processing}

Signal processing operations, such as encoding/decoding, interleaving/deinterleaving, modulation/demodulation, and scrambling/descrambling, are performed in this block. 
Encoding for the channel serves as an error correcting code. 
Specifically, the encoded signal includes redundancy that is utilized by the receiver's decoder to re-construct the original signal from the corrupted received signal. 
Examples of error correcting codes include Convolutional Codes, Turbo Codes, and Low Density Parity Check (LDPC) \cite{choo14}. 
The decoder constitutes the most computationally intensive part of the Signal Processing block, due to data transfer and memory schemes \cite{Berns}. 
The second part that is regarded as highly complex and expensive (in terms of area and power) is the Fast Fourier Transform (FFT) and Inverse FFT (IFFT), as part of the modulation phase \cite{ChiuehTzi-DarandTsai2008}. 

The signal processing block is commonly referred to as the \textit{baseband processing block}. 
When discussing SDRs, the baseband block is at the heart of the discussion, as it makes up the bulk of the digital domain of the implementation. 
This implementation runs on top of a hardware circuitry that is capable of processing signals efficiently. 
Examples include ASICs, FPGAs, DSPs, GPPs, and GPUs. 
The second part of the implementation is the software, which provides the functionality and high-level abstractions to execute signal processing operations. 
In the next section we examine the aforementioned hardware platforms and analyze in detail the various design approaches.

\section{Design Approaches}
\label{hardware}

In this section, we discuss the classification of the various SDR design methodologies of the baseband processing block, namely GPP, GPU, DSP, FPGA, and co-design based methodologies. 
In this classification, we analyze and compare SDR platforms based on a set of performance metrics in a criteria which we introduce. 
This criteria includes: 
\begin{itemize}
\renewcommand\labelitemi{--}
\item \textit{Flexibility and Reconfigurability}. The capability for the modulation and air-interface algorithms and protocols to evolve by merely loading new software onto the platform \cite{Angeletti2014}.
\item \textit{Adaptability}. The SDR platform can adjust its capabilities as network or traffic operational requirements change.
\item \textit{Computational Power}. The processing rate of the SDR platform, namely Giga Operations per Second (GOPS).
\item \textit{Energy Efficiency}. The total power consumption (typically within a few hundreds milli watts), especially for mobile \cite{Dezfouli2017} and IoT deployments. 
\item \textit{Cost}. The total cost of the SDR platform, including time-to-market, development, and hardware costs.\\
\end{itemize}

\subsection{GPP-based}
\label{gpp_sec}

One of the first approaches to realizing SDR platforms is using a \ac{gpp}, or the commonly known generic computer microprocessors such as x86/64 and ARM architectures. 
Examples of SDR platforms that utilize GPPs include Sora \cite{Tan2011}, KUAR \cite{kuar2007}, and USRP \cite{usrp}.

\subsubsection{Definition and Uses}

A \ac{gpp} is a digital circuit that is clock-driven and register-based, and is capable of processing different functions and operates on data streams represented in binary system \cite{kant2014}. 
These GPPs can be used for several purposes, making them extremely useful for unlimited number of applications, eliminating the need for building application-specific circuits, and thus reducing the overall cost of running applications. 
GPPs are generally a preferable hardware platform by researchers in academia due to their flexibility, abundance, and ease of programmability, which is one of the main requirements in SDR platforms \cite{Kazaz2017}. 
In addition, researchers prefer GPPs since they are more familiar with them and their software frameworks, compared to DSPs and FPGAs.
From the performance point of view, GPPs are being enhanced rapidly, credited not only to technological advances in terms of CMOS technology \cite{cmos2014}, but also to the increase of the average number of instructions processed per clock cycle. 
The latter is achieved through different means, and in particular, utilizing parallelism within and between processors. 
This has led to the evolution of multi-core GPPs \cite{Ulversoy2010}.

\subsubsection{Adoption and GPUs}
\label{gpu}

Architecturally, the instruction set of GPPs include instructions for different operations such as Arithmetic and Logic Unit (ALU), data transfer, and I/O. 
A GPP processes these instructions in the sequential order. 
Because of sequential processing, GPPs are not convenient for high-throughput computing with real-time requirements (i.e., high throughput and low latency) \cite{Kamal2003}. 
For example, using GNU Radio \cite{gnu} to implement IEEE 802.11 standard, which requires 20MHz sampling rate, would be challenging, since GNU Radio is restricted by the limited processing capabilities of GPPs. 
This leads to the GPP cores (of the PC attached) to reach saturation and frames become corrupted and discarded.
Moreover, wireless protocols require predictable performance in order to guarantee meeting timing constraints.
However, conditional branch instructions in GPP's instruction sets lead to out-of-order execution, which makes it unfeasible to achieve predictability.

To remedy the limitation of GPPs, researchers have proposed multiple solutions, one of which is the addition of co-processors, such as Graphic Processing Unit (GPU) \cite{Vachhani2015}. 
GPUs are processors specifically designed to handle graphics-related tasks and they efficiently process large blocks of streaming data in parallel.
SDR platforms comprised of both GPPs and GPUs are flexible and have a higher level of processing power.
However, this results in a lower level of power efficiency (e.g., GPP's power efficiency is $\sim$9GFLOPS/W for single-precision, compared to 20GFLOPS/W for GPU \cite{Vestias2014}).
GPUs act as co-processors to GPPs because a GPP is required to act as the control unit and transfer data from external memory. 
After a transfer is completed, signal processing algorithms are executed by the GPU. 


While GPUs are typically used for processing graphics, they are also useful at signal processing algorithms. 
Over the past few years, theoretical peak performance for GPUs and GPPs for single and double precision processing has been growing \cite{cpu_gpu}. 
For example, comparing Intel Haswell's 900 GFLOPs \cite{intel} with NVIDIA GTX TITAN's 4500 GFLOPS \cite{nvidia} for single precision, it is apparent that GPUs have a computational power that far exceeds their GPP counterparts \cite{cpu_gpu}. 
Their multi-core architectures and parallel processors are the main attractive features, in addition to their relatively reasonable prices and credit card sizes. 
These features makes them good candidates as co-processors in GPP-based SDRs, where they can play a vital role in accelerating computing-intensive blocks \cite{Fisne2017}. 
Another advantage is their power efficiency, which keeps improving with every new model (e.g., it went from 0.5 to 20GFLOPS/W for single-precision) \cite{Vestias2014}. To take full advantage of GPUs, it is a condition that algorithms conform to their architecture. 
From an architectural perspective, GPUs have a number of advantages that makes them preferable solutions to applications such as video processing. 
In particular, GPUs employ a concept called Single Program Multiple Data (SPMD) that allows multiple instruction streams to execute the same program. 
In addition, due to their multi-threading scheme, data load instructions are more efficient. 
GPUs also present a high computational density, where cache to ALU ratio is low \cite{Cope2010}.

In Table \ref{cpu_gpu_comp}, the authors of \cite{Fisne2017} confirmed that the signal detection algorithm (which includes intensive FFT computations) shows a faster parallel processing in the case of GPU over GPP, while operating in real-time. 
This is due to the availability of cuFFT library developed for NVIDIA GPUs for more efficient FFT processing \cite{cuda}. 
With regards to the architectural advantage of GPUs, several hundred CUDA cores can perform a single operation at the same time, as opposed to a few cores in the case of multi-core GPPs. 


\begin{table*}[]
\centering
\caption{Performance of Signal Detection Algorithm on GPP and GPU \cite{Fisne2017}}
\label{cpu_gpu_comp}
\begin{tabular}{|c|c|c|c|}
\hline & \multicolumn{3}{c|}{Processing Platform of Signal Detection Algorithm} \\ \cline{2-4}
\multirow{-2}{*}{\begin{tabular}[c]{@{}c@{}}\textbf{ADC Data Length (ms)}\end{tabular}} & \textbf{GPP Serial Processing (ms)}  & \textbf{GPP Parallel Processing (ms)} & \textbf{GPU Parallel Processing (ms)} \\ \hline
1                                                                                                        & 13.487                      & 1.254                        & 0.278                        \\ \hline
10                                                                                                       & 135.852                     & 12.842                       & 2.846                        \\ \hline
100                                                                                                      & 1384.237                    & 131.026                      & 29.358                       \\ \hline
1000                                                                                                     & 13946.218                   & 1324.346                     & 321.254                      \\ \hline
\end{tabular}
\end{table*}

Examples of using GPUs alongside GPPs to build SDR platforms is the work in \cite{gpu2009}, where the authors built a framework on a desktop PC in addition to using a GPU to implement an FM receiver. 
The authors in \cite{Fisne2017} studied real-time signal detection using an SDR platform composed of a laptop computer and an NVIDIA Quadro M4000M \cite{nvidia}.
Examples of GPUs available in the market can be found in Table \ref{gpu_comp}.
In this table, we show two examples of high performing GPUs ($>5500$ GFLOPS) suitable for SDRs with strict timing and performance requirements. 
We also show two more examples of less powerful and less expensive GPUs suitable for prototyping SDRs in academic environments.

\begin{table*}[!th]
\centering
\scriptsize
\renewcommand{\arraystretch}{1.1}
\setlength{\tabcolsep}{0.7em} 
\caption{Comparison of GPUs}
\label{gpu_comp}
\begin{tabular}{c|c|c|c|c|}
\cline{2-5}
                                                     & \textbf{\begin{tabular}[c]{@{}c@{}}NVIDIA GeForce\\ GTX 980 Ti \cite{nvidia}\end{tabular}} & \textbf{\begin{tabular}[c]{@{}c@{}}AMD Radeon\\ R9 390X \cite{AMD}\end{tabular}} & \textbf{\begin{tabular}[c]{@{}c@{}}NVIDIA  GeForce\\ GTX 680 \cite{nvidia}\end{tabular}} & \textbf{\begin{tabular}[c]{@{}c@{}}AMD Radeon\\ RX 560 \cite{AMD}\end{tabular}} \\ \hline
\multicolumn{1}{|c|}{\textbf{GFLOPS}}                & 5632                                                                         & 5913                                                                  & 3090                                                                       & 2611                                                                 \\ \hline
\multicolumn{1}{|c|}{\textbf{Power Consumption (W)}} & 250                                                                          & 363                                                                   & 356                                                                        & 180                                                                  \\ \hline
\multicolumn{1}{|c|}{\textbf{Frequency (MHz)}}       & 1000                                                                         & 1050                                                                  & 1006                                                                       & 1175                                                                 \\ \hline
\multicolumn{1}{|c|}{\textbf{Cost (USD)}}            & 870                                                                          & 520                                                                   & 300                                                                        & 150                                                                  \\ \hline
\end{tabular}
\end{table*}

\subsubsection{Shortcomings}

State-of-the-art GPP and GPU-based platforms, such as Sora \cite{Tan2011} and USRP \cite{usrp}, utilize desktop computers to realize the systems. 
However, these platforms consume a significant amount of power for a performance goal and their form factor is large, which makes it impossible for real-world deployments. 
It is worth noting that GPPs and GPUs alike, present scaling limitation while meeting Koomey's Law.
This law states that energy efficiency of computers doubles roughly every 18 months \cite{koomey2011}. 
This limitation calls for alternatives that provide higher computing power while keeping the energy efficiency the same. 
One alternative is the \textit{hybrid} or \textit{co-design }approach, where software and hardware implementations are combined. 
This will be discussed in more details in Section \ref{hybrid}.

When both GPP and GPU are used for a SDR design, data transfer operations between GPP and GPU can be bottlenecks and cause performance loss, especially for meeting real-time requirements \cite{Li2014}. 
However, there are continuous efforts to reduce or eliminate the time overhead of data transfers by introducing multi-stream scheduling for pipelining of the memory copy tasks. 
This would ensure no stalls in the pipeline and thus enhancing processing parallelism \cite{Li2015,Millage2010}. 
Finally, although the processing power of microprocessors is being constantly improved, balancing between sufficient computing power and a specific goal for energy consumption and cost, stays a very difficult task now and in the future. 
This is true especially with the growing need for more data to be processed and blocks that can handle data processing in parallel.

\subsection{DSP-based}

The DSP-based solution can be considered as a special case of GPP-based solutions, but due to its popularity and unique processing features, it deserve a separate discussion. 
An example of DSP-based SDR is the Atomix platform \cite{Bansal2015} which utilizes TI TMS320C6670 DSP \cite{tia}.

\subsubsection{Definition and Uses}

DSP is a particular type of microprocessor that is optimized to process digital signals \cite{Rabiner1975}. 
To help understand how DSPs are distinguished from GPPs, we should first note that both are capable of implementing and processing complex arithmetic tasks \cite{Smith1997}. 
Tasks like modulation/demodulation, filtering, and encoding/decoding are commonly and frequently used in applications that include speech recognition, image processing, and communication systems. 
DSPs, however, implement them more quickly and efficiently due to their architecture (e.g., RISC-like architecture, parallel processing) which is specifically optimized to handle arithmetic operations, especially multiplications. 
Since DSPs are capable of delivering high performance with lower power, they are better candidates for SDR deployment \cite{dsp93}, compared to GPPs.
Examples of DSPs especially designed for SDR platforms are TI TMS320C6657 and TMS320C6655.
These DSPs are both equipped with hardware accelerators for complex functions like the Viterbi and Turbo Decoders \cite{ti}.

\subsubsection{Adoption}

As discussed in the previous section, GPPs provide an average performance for a wide range of applications. 
Needless to say, this performance level might be sufficient for research and academia, but if the system is to be deployed commercially, certain performance requirements must be met. 
To this end, compared to GPPs, DSPs are tailored for processing digital signals efficiently, utilizing features like combined multiply-accumulate operations (MAC units) and parallelism \cite{Antoniou2016}. 
DSP manufacturers usually sell these products in two flavors: optimized for performance, and energy optimized.
Therefore, when used in SDRs, high performance and energy efficient products can be employed in BSs and edge devices, respectively.

In terms of the instruction set, DSPs can be categorized into two groups: 
(i) Single Instruction Multiple Data (SIMD) architecture, and 
(ii) Multiple Instruction Multiple Data (MIMD) architecture, as described by Michael J. Flynn in what is known as Flynn's Taxonomy \cite{patterson,flynn72}. 
This taxonomy is a method of classifying various architectures depending on the number of concurrent instructions and data streams, as follows:%
\begin{itemize}
\renewcommand\labelitemi{--}
\item A SIMD-based DSP can execute an instruction on multiple data streams at the same time. 
This architecture can be very efficient in cases when there exists high data parallelism within the algorithm \cite{duncan90}. 
This indicates that there are similar operations that can be performed on different datasets at the same time. 
Examples of SIMD-based DSPs include the Cell processor presented in \cite{cell2006} which supports 256 GFLOPS. 
More examples of DSPs that are optimized for low power are NXP CoolFlux DSP \cite{nxp} and Icera Livanto \cite{icera}. 
A SDR employing a SIMD DSP is the SODA architecture \cite{Lin2006}. 
It has been a common practice to add more cores in order to achieve a better trade off between performance and power. 
With each extra core utilizing Very Long Instruction Word (VLIW), a higher level of parallelism can be accomplished as well. 
\item On the other hand, MIMDs have the ability to operate on multiple data streams executing multiple instructions at any point in time. 
This is essentially an extension of the SIMD architecture, where different instructions or programs run on multiple cores concurrently. 
This is especially important and useful in cases where parallelism is not uniform across different blocks, but MIMD architecture allows for parallel execution which leads to speed improvements. Examples of MIMD-based DSPs include Texas Instruments SMJ320C80 and SM320C80 DSPs with 100 MFLOPS \cite{ti}. 
\end{itemize}

Since DSPs are customized to meet certain signal processing-related needs, it is crucial to clarify these customizations in order to understand how DSPs stand out and how they are successful at not only meeting the requirements, but also becoming a vital player in the wireless communication field. 
These customizations, which are mostly architecture-related, are as follows.

In \cite{Gatherer2000}, the authors discuss the energy efficiency of DSPs. 
In general, DSPs consume more power than ASICs, however, there exists DSPs that are optimized for low power wireless implementations, such as TI C674x DSP \cite{ti}.
One of the methods to lower power consumption is using multiple data memory buses (e.g., one for write, and two for reads). 
This paves the way for higher memory bandwidth, and allows for multiple operand instructions, which in turn results in fewer cycles. 
Also, as discussed above, VLIW architectures along with specialized instructions can provide a higher level of efficiency, and hence lower energy consumption. 
These improvements can be seen in DSPs such as TI TMS320C6x \cite{ti} and ADI TigerSHARC \cite{AD}. 
These techniques, coupled with proven power saving techniques like clock gating and putting parts of or the entire system in sleep mode, further reduce the power consumption.
Examples of DSPs available in the market can be found in Table \ref{dsp_comp}. 
In this table, we present three examples of DSPs that do no include co-processors, and three DSP-based SoCs that, in addition to DSP cores, include extra soft cores as control processors. 

\begin{table*}[!ht]
\centering
\scriptsize
\caption{Comparison of DSPs and DSP-based SoCs}
\label{dsp_comp}
\begin{tabular}{c|c|c|c|c|c|c|}
\cline{2-7}
                                                                       & \multicolumn{3}{c|}{\textbf{DSP only}}                                                                                                                                                                                               & \multicolumn{3}{c|}{\textbf{SoC}}                                                                                                                                                                                                                                                                 \\ \cline{2-7} 
                                                                       & \textbf{\begin{tabular}[c]{@{}c@{}}TI C66x \\ (TMS320C6652) \cite{ti}\end{tabular}} & \textbf{\begin{tabular}[c]{@{}c@{}}CEVA\\ (XC-4500)\cite{ceva}\end{tabular}} & \textbf{\begin{tabular}[c]{@{}c@{}}Analog Devices\\ (ADSP-21369)\cite{AD}\end{tabular}} & \textbf{\begin{tabular}[c]{@{}c@{}}TI Keystone II\\ (66AK2G02) \cite{ti}\end{tabular}} & \textbf{\begin{tabular}[c]{@{}c@{}}Analog Devices\\ (ADSP-SC573)\cite{AD}\end{tabular}} & \textbf{\begin{tabular}[c]{@{}c@{}}Qualcomm Snapdragon 820 \\ (Hexagon 680) \cite{qualcomm}\end{tabular}} \\ \hline
\multicolumn{1}{|c|}{\textbf{GFLOPS}}          & 9.6                                                                                                  & 40                                           & 2.4                                                                                                         & 28.8                                                                                                 & 5.4 & No Floating Point                                                                                                           \\ \hline
\multicolumn{1}{|c|}{\textbf{Memory (Kb)}}     & 1088                                                                                                 & No Info                                            & 2000 & 1024                                                                                                 & 768                                                                                                        & No Info \\ \hline
\multicolumn{1}{|c|}{\textbf{Frequency (MHz)}} & 600                                                                                                  & 1300                                         & 400                                                                                                        & 600                                                                                                  & 450                                                                                                        & 2000                                                                                                            \\ \hline
\multicolumn{1}{|c|}{\textbf{Cost (USD)}}      & $\sim$25                                                                                                 & No Info                                             & $\sim$20                                                                                                        & $\sim$20                                                                                                & $\sim$20                                                                                                        &$\sim$70 \\ \hline
\multicolumn{1}{|c|}{\textbf{Soft Core}}       & N/A                                                                                                    & N/A                                            & N/A                                                                                                      & \begin{tabular}[c]{@{}c@{}}ARM \\ Cortex-A15\end{tabular}                                            & \begin{tabular}[c]{@{}c@{}}ARM \\ Cortex-A5\end{tabular}                                                                                                         & \begin{tabular}[c]{@{}c@{}}Qualcomm Kyro 385 (CPU)\\ Adreno 530 (GPU)\end{tabular}                                                                                                                                                                                                                   \\ \hline
\end{tabular}
\end{table*}

\subsubsection{Shortcomings}

Although DSPs have been widely adopted for SDR implementations for decades \cite{Frantz2000}, they present shortcomings as follows. 
As more applications call for increasing parallelism and reconfigurability in order to handle computationally intensive tasks, DSPs can be insufficient. 
In addition, programming DSPs to achieve higher levels of parallelism predictability can be challenging.
This opened the door for parallel architectures like FPGAs, or multi-core GPPs, or even a hybrid of both, to be adopted for SDRs. 
In addition, power consumption of DSPs is generally higher than FPGAs due to operating at high frequencies.

\subsection{FPGA-based}
\label{fpga_method}

Another approach towards realizing SDRs is to use a programmable hardware such as FPGAs. 
Example of FPGA-based SDR platforms are Airblue \cite{Ng2010}, Xilinx Zynq-based implementation of IEEE 802.11ah \cite{Akeela2017}, and \cite{Cai2017b} that used the same FPGA board to implement a complete communication system with channel coding.

\subsubsection{Definition and Uses}

An FPGA is an array of programmable logic blocks, such as general logic, memory, and multiplier blocks, that are surrounded by a routing fabric, which is also programmable \cite{fpga}. 
This circuit has the capability of implementing any design or function, with ease of updating it.
Although FPGAs consume more power and occupy more area than ASICs, the programmability feature is the reason behind their increasing adoption in a wide range of applications.
Furthermore, when the reconfiguration delay is in the order of milliseconds, the SDR can switch between different modes and protocols seamlessly \cite{Grover2012}. 
Another major difference is that, ASIC fabrication is expensive (at least a few tens of thousands of dollars) and requires a few months, whereas FPGAs can be quickly reprogrammed, and their cost is within a few tens to a few thousands of dollars, at most. 
The low end product cycle, along with attractive hardware processing advantages, such as high speed performance, low power consumption, and portability, compared to processors such as GPPs and DSPs, present FPGAs as contenders that offer the best of both worlds \cite{fpga}.

\begin{figure}[t]
\centering
\includegraphics[width=0.9\linewidth]{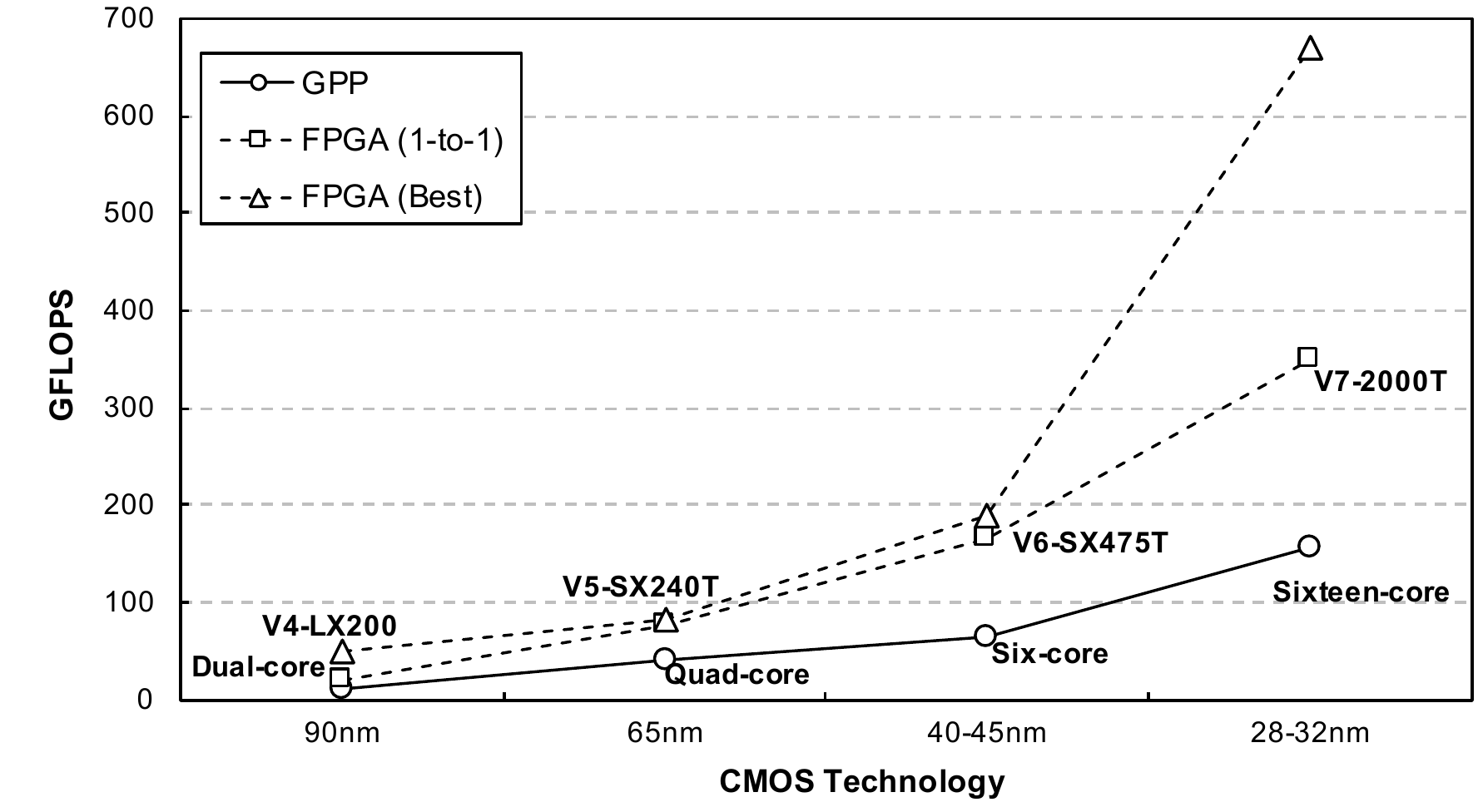}
\caption{Peak performance of GPPs versus FPGAs when performing 64-bit floating point operations \cite{Strenski2012}. It can be observed that FPGAs increased their floating point performance by an order of magnitude compared to GPPs.}
\label{cpu_vs_fpga}
\end{figure}

In a study by \cite{Strenski2012}, the authors compared the performance of Xilinx FPGAs \cite{xilinx} against 16-core GPPs. 
The calculation of peak performance for GPPs was performed through multiplying the number of floating point function units on each core by the number of cores and by the clock frequency.
For FPGAs, performance is calculated through picking a configuration, adding up the Lookup Tables (LUTs), flip-flops, and DSP slices needed, then multiplying them by the appropriate clock frequency. 
The authors calculated the theoretical peaks for 64-bit floating point arithmetic and showed that Xilinx Virtex-7 FPGA is about 4.2 times faster than a 16-core GPP. 
This can been seen in Figure \ref{cpu_vs_fpga}. 
Even with a one-to-one adder/multiplier configuration, the V7-2000T achieved 345.35GFLOPS, which is better than a 16-core GPP.
From Intel \cite{intel}, Stratix 10 FPGAs can achieve a 10 Tera FLOPS peak floating point performance \cite{Lewis2016}. 
This is due to the fixed architecture of the GPP, where not all functional units can be fully utilized, and the inherent parallelism of FPGAs and their dynamic architecture. 
In addition, despite having lower clock frequencies (up to 300MHz), FPGAs can achieve better performances due to their architectures which allow higher levels of parallelism through custom design \cite{Sano2017}. 
In a study by \cite{Kestur2010}, the authors compared the performance and power efficiency of FPGAs to that of GPPs and GPUs using double-precision floating point matrix-vector multiplication. 
The results show that FPGAs are capable of outperforming the other platforms, while maintaining their flexibility. 
In another study by \cite{Cope2010}, the authors thoroughly analyzed and compared FPGAs against GPUs via the implementations of various algorithms.
The authors concluded that although both architectures support a high level of parallelism, which is crucial to signal processing applications, FPGAs offer a larger increase in parallelism, whereas GPUs have a fixed parallelism due to their data path and memory system.

\subsubsection{Adoption}

Over the past decade, FPGAs have significantly advanced and become more powerful computationally, and now exist in many differnet versions such as Xilinx Kintex UltraScale \cite{xilinx} and Intel Arria 10 \cite{intel} \cite{dspfpga2008,dspfpga2014}. 
In addition, the availability of various toolsets gave FPGAs an advantage by making them more accessible. 
This is supported by the availability of compilers that have the capability of generating Register-transfer Level (RTL) code, such as Verilog and VHDL, that is needed to run on FPGAs, from high-level programming languages. 
This process is typically referred to as \textit{High Level Synthesis} (HLS). 
Examples of such compilers include HDL Coder \cite{hdl_coder} for MATLAB code \cite{matlab} and Xilinx HLS \cite{hls} or Altera Nios II C2H compiler \cite{altera} for C, C++, and SystemC. We will explain some of these tools in Section \ref{tools}. 

HLS allows software engineers to design and implement applications, such as SDRs, on FPGAs using a familiar programming language to code, namely C, C++, SystemC, and MATLAB, without the need to posses a prior rich knowledge about the target hardware architecture (refer to Section \ref{hls}). 
These compilers can also be used to speed up or accelerate parts of the software code running on a GPP or DSP that are causing slowdowns or setbacks to the overall performance. This will be further discussed in Section \ref{hybrid}. 
Further, FPGAs can achieve high performance while still consuming less energy than previously discussed processors \cite{dspfpga2003} (e.g., Intel Stratix 10 FPGA can achieve up to 100 GFLOPS/W \cite{Corporation}, compared to 23 GFLOPS/W for NVIDIA GeForce GTX 980 Ti \cite{nvidia}). 
In addition, power dissipation can be further lowered through the implementation of several techniques discussed in \cite{Grover2012}. 
These techniques can be at a system, device, and/or architecture level, such as clock gating and glitch reduction. 
Table \ref{fpga_comp} presents a summary of widely-used FPGA platforms.

\subsubsection{Shortcomings}

One of the challenges of using FPGAs, however, can be the prior knowledge about target hardware architecture and resources that a developer needs to possess in order to design an application efficiently for FPGAs. In the SDR domain, designing the platform has typically been the job of software engineers, and thus the process can be time-consuming and less trivial to incorporate this experience into hardware design. However, as it will be discussed in Section \ref{hls}, adoption of FPGA solutions can be made more feasible through HLS tools. 

\begin{table*}[]
\scriptsize
\centering
\caption{Comparison of FPGAs and FPGA-based SoCs}
\label{fpga_comp}
\begin{tabular}{c|c|c|c|c|c|c|}
\cline{2-7}
                                                                       & \multicolumn{3}{c|}{\textbf{FPGA only}}                                                                                                                                                                                                                                                    & \multicolumn{3}{c|}{\textbf{SoC}}                                                                                                                                                                                                                                                                        \\ \cline{2-7} 
                                                                       & \textbf{\begin{tabular}[c]{@{}c@{}}Xilinx Kintex-7\\ (XC7K70T) \cite{xilinx}\end{tabular}} & \textbf{\begin{tabular}[c]{@{}c@{}}Intel Cyclone V GX\\ (C5) \cite{intel}\end{tabular}} & \textbf{\begin{tabular}[c]{@{}c@{}}Lattice ECP3-70\\ (LFE3-70EA) \cite{lattice}\end{tabular}} & \textbf{\begin{tabular}[c]{@{}c@{}}Xilinx Zynq-700\\ (Z-7020 XC7Z020) \cite{xilinx}\end{tabular}} & \textbf{\begin{tabular}[c]{@{}c@{}}Intel Cyclone V SE SoC\\ (A5) \cite{intel}\end{tabular}} & \textbf{\begin{tabular}[c]{@{}c@{}}Microsemi SmartFuion2\\ (M2S090) \cite{microsemi}\end{tabular}} \\ \hline 
\multicolumn{1}{|c|}{\textbf{Logic Cells (K)}} & 65.6                                                                                                 & 77                                                                                                 & 67                                                                                                     & 85                                                                                                          & 85                                                                                                     & 86.31                                                                                                     \\ \hline
\multicolumn{1}{|c|}{\textbf{Memory (Mb)}}     & 4.86                                                                                                 & 4.46                                                                                               & 4.42                                                                                                   & 4.9                                                                                                         & 3.97                                                                                                   & 4.488                                                                                                     \\ \hline
\multicolumn{1}{|c|}{\textbf{DSP Slices}}      & 240                                                                                                  & 150                                                                                                & 128                                                                                                    & 220                                                                                                         & 87                                                                                                     & 84                                                                                                        \\ \hline
\multicolumn{1}{|c|}{\textbf{Cost (USD)}}      & 130 & 185                                                                                                & 80 & 110 & 110 & 155 \\ \hline
\multicolumn{1}{|c|}{\textbf{Soft Core}}       & N/A                                                                                                    & N/A & N/A & \begin{tabular}[c]{@{}c@{}}Dual-core ARM \\ Cortex-A9\end{tabular}                                          & \begin{tabular}[c]{@{}c@{}}Dual-core ARM\\ Cortex-A9\end{tabular}                                      & ARM Cortex-M3                                                                                             \\ \hline
\end{tabular}
\end{table*}

\subsection{Hybrid Design (a.k.a., co-design)}
\label{hybrid}

The fourth approach towards realizing SDRs is the hybrid approach, where both hardware and software-based techniques are combined into one platform. 
This is commonly referred to as the \textit{co-design} or \textit{hybrid} approach. 
Examples of SDRs that adopted the co-design approach include WARP \cite{warp} and CODIPHY \cite{Dutta2013}.

\subsubsection{Definition}

Hardware/software co-design as a concept has been around for over a decade, and it has evolved at a faster rate in the past few years due to an increasing interest in solving integrated circuit design problems with a new and different approach. 
Even with GPPs becoming more powerful than ever, and with multi-core designs, it is clear that in order to achieve higher performance and realize applications that demand real-time processing, designers had to shift attention to new design schemes that utilize hardware solutions, namely, FPGAs and ASICs \cite{wolf2003,micheli2002}. Co-design indicates the use of hardware design methodology, represented by the FPGA fabric, and software methodology, represented by processors. 

As more applications, such as automotive, communication, and medical, grow in complexity and size, it has become a common practice to design systems that integrate both software (like firmware and operating system) and hardware \cite{Teich2012}. 
This has been made feasible in the recent years thanks to the advances in high-level synthesis and developing tools that not only have the capability to produce efficient RTL from software codes, but also define the interface between the both sides. 
The industry has realized the huge market for co-design, and provided various SoC boards, that in addition to the FPGA fabric, contain multiple processors. 
For example, the Xilinx Zynq board \cite{xilinx} includes an FPGA fabric as well as two ARM Cortex-A9 processors \cite{arm}.
In addition to the aforementioned advantages, there are other reasons that make co-design even more interesting including, faster time-to-market, lower power consumption (when optimized for this), flexibility, and higher processing speeds, as typically hardware in these systems is used as an acceleration to software bottlenecks \cite{Windh2015}. 


Adopting the co-design methodology in essence is a matter of partitioning the system into synthesizable hardware and executable software blocks. 
This process depends on a strict criteria that is developed by the designer \cite{Chehida2002,Lopez-Vallejo2003}. 
The authors in \cite{Zhuo2007} and \cite{Sapienza2014} discuss their partitioning methodologies and present the process of making the proper architectural decisions. Common methods typically provide useful information to the designer to help make the best decision of what to implement in hardware and what to keep in software. 
This information can include possible speedups, communication overheads, data dependencies, and locality and regularity of computations \cite{Sapienza2014}. 
Examples of SoC boards available in the market can be found in Table \ref{fpga_comp}.


\subsubsection{Adoption}

As mentioned in Section \ref{arch}, SDRs can be considered inherently hybrid or heterogeneous systems, thereby implying the need for both hardware and software blocks. This is due to the fact that the control part is usually taken care of by a general processor, and other functions, such as signal processing, by a specialized processor like DSPs, and sometimes are accelerated using dedicated hardware like FPGAs \cite{bolsens97}. 
This design approach fits well with SDRs and can be fully utilized to meet certain requirements that pertain to their attractive features. 
For example, accelerating portions of a block or moving it entirely to the FPGA fabric can help to push the processing time to the limit in order to achieve a real-time performance for real-life deployment. 
In addition, through careful implementation of RTL optimization techniques, the development of power efficient systems for mobile and IoT applications would be possible. 
On the other hand, running most of the MAC layer operations on a processor, or multi-processors, can be advantageous for easy reconfiguration. 
Therefore, different partitioning schemes can be adopted to meet the requirements of the application at hand. 

\begin{figure}[!t]
\centering
\includegraphics[width=0.9\linewidth]{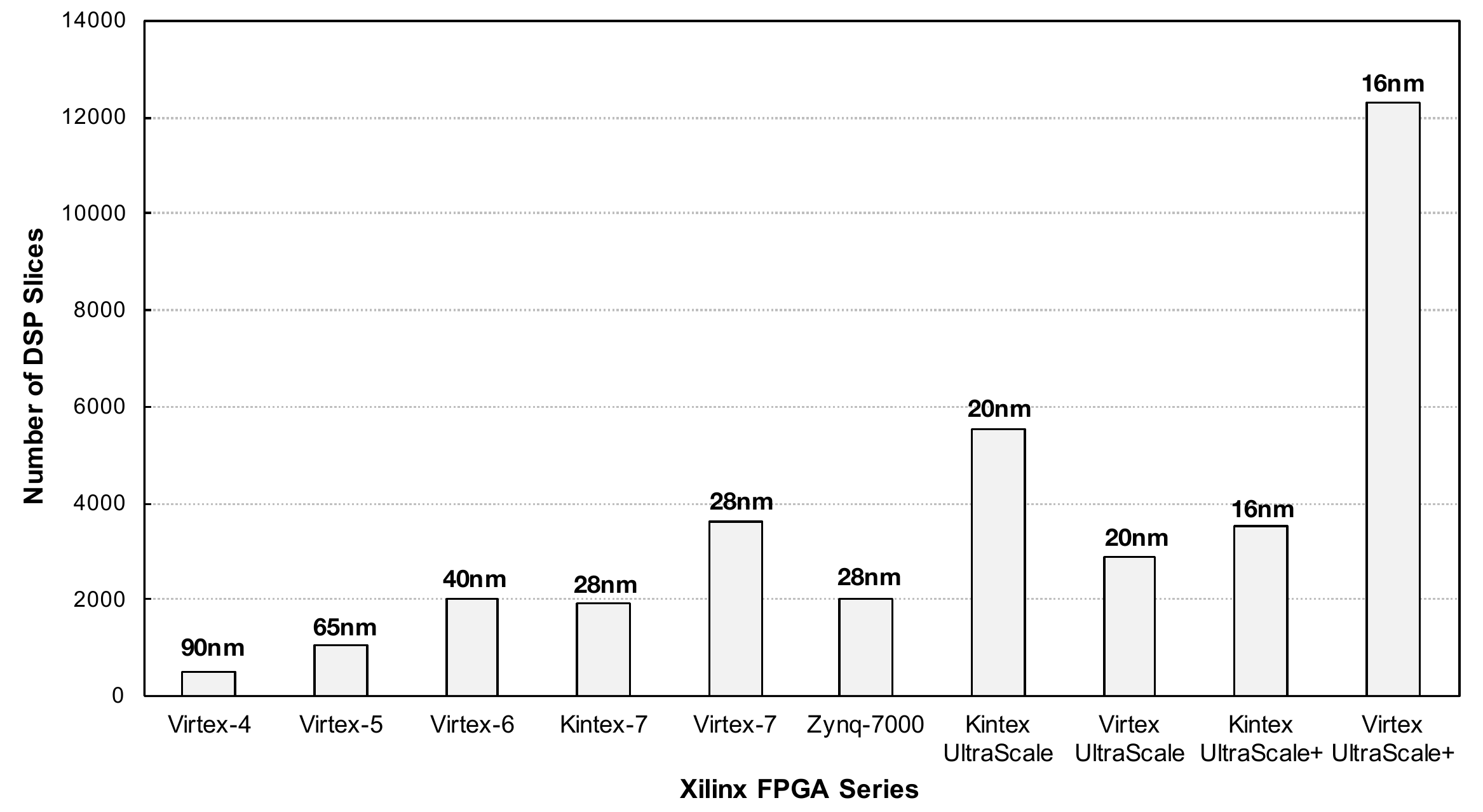}
\caption{Number of DSP Slices in Xilinx FPGAs \cite{xilinx}. The values on top of the bars refer to the CMOS technology used.}
\label{dsp_x}
\end{figure}

It is worth noting that FPGA vendors, namely Intel \cite{intel} and Xilinx \cite{xilinx}, are widening their product base with more SoCs and Multi-processor SoCs (MPSoCs) \cite{Zhang2015a}, due to the the growing demand for such devices. 
An example of an SDR realized on an MPSoC is the work by \cite{Halim2015}. In a white paper, National Instruments (the company that owns USRP \cite{usrp}) predicts that the future of SDRs is essentially a co-design implementation \cite{sni04}, especially with the introduction of FPGAs that are equipped with a large number of DSP slices for handling intensive signal processing tasks, as depicted in Figure \ref{dsp_x}. 
This also can be seen from USRP E310 model, which incorporates a Xilinx Zynq SoC \cite{xilinx}.

\subsubsection{Shortcomings}

A downside of adopting SoCs for co-design is that their prices are generally higher, compared to previously mentioned design approaches, due to having multiple components on the same board, i.e., processor and FPGA fabric. 
Other factors contributing to this include extra memory and sophisticated interfaces. 
Another challenge of co-design is shared memory access, e.g., external DDR memory, between the processor and FPGA fabric. 
The study of \cite{Gobel2017} shows that the number of memory read and write operations performed by a GPP is higher than that of FPGAs.
This is due to the fact that processors perform operation on registers, while FPGAs operate on buffers.
Since memory accesses add up to the overall latency, this can cause a bottleneck to the overall performance. 
In addition, the authors have developed a methodology for predicting shared memory bandwidth by using a functionally-equivalent software. 
This enables the designers to be aware of any bottlenecks before implementing the entire co-design.

\subsection{Comparison}
\label{comp}

When we covered different design methodologies and hardware platforms for a wide selection of SDR platforms, we intended to compare them analytically one-on-one, using a cross-platform implementation of one of the wireless communication protocols. What was available in literature, instead, was a series of abstract comparisons using a set of benchmarks targeting High Performance Computing (HPC), and not necessarily SDR applications. It is somewhat difficult to draw conclusions from these numbers alone, since a performance comparison in the SDR field requires testing them in real-life.

In Table \ref{sdr_comp}, we provide a high-level comparison between three major design approaches as a guideline for designers towards choosing the method that best meets their application specifications. 
In this comparison, we focus on the features that are important to SDR design. 
However, we do not make assumptions on what the best approach is and believe it is the developer's responsibility to make the best judgment depending on the application area. 
Please note that in this table we did not include GPUs, as they typically act as co-processors to GPPs and their addition generally improves performance. 
We also did not include co-design since it combines GPPs with FPGAs.

As Table \ref{sdr_comp} shows, while GPPs are easy to program and extremely flexible, they lack the power to meet specifications in real-time and are very inefficient in terms of power. 
To increase their performance, multiple cores with similar instruction sets are included in the same GPP platform to exploit parallelism and perform more operations per clock cycle. 
However, hardware replication (i.e., adding more cores to GPPs) may not necessarily translate to a higher performance. 
GPUs tackle this by offering the same control logic for several functional units. 
The sequential portion of the code runs on the GPP, which can be optimized on multi-core GPPs, while the computationally intensive portion runs on a several-hundred-core GPU, where the cores operate in parallel.
Another example of a customized processor is DSP, which performs significantly better than GPPs, while at the same time maintaining the ease-of-use feature that GPPs possess, making them very attractive options. 
They are also more power efficient and better fit for signal processing applications. 
On the other hand, they are more expensive, which is the main trade-off. 
Finally, FPGAs combine the flexibility of processors and efficiency of hardware. 
FPGAs can achieve a high level of parallelism through dynamic reconfiguration, while yielding better power efficiency \cite{Vestias2014}. 
FPGAs are typically more suitable for fixed-point arithmetic, like signal processing tasks, but in the recent years their floating-point performance has increased significantly \cite{Underwood2004,Kestur2010}. 
However, the designers are expected to know a lot more about the hardware, which is sometimes a deterring feature.

In a comparative analysis by \cite{Betkaoui2010}, authors studied the performance and energy efficiency of GPUs and FPGAs using a number of benchmarks in terms of targeted applications, complexity, and data type. 
The authors concluded that GPUs perform better for streaming applications, whereas FPGAs are more suitable for applications that employ intensive FFT computations, due to their ability to handle non-sequential memory accesses in a faster and more energy efficient manner. 
Similarly, in \cite{Vestias2014}, the authors review and report the sustainable performance and energy efficiency for different applications. 
One of their findings related to SDRs is that FPGAs should be used for signal processing without floating point, confirming aforementioned results. 
In addition, the authors in \cite{duan2011} report that GPUs are ten times faster than FPGAs with regards to FFT processing, while authors in \cite{Kestur2010} demonstrate that the power efficiency of FPGAs is always better than GPUs for matrix operations.
Finally, authors in \cite{Owaida2015} compare GPPs, GPUs, and FPGAs through the implementation of LDPC decoders, and their results lead to the conclusion that GPUs and FPGAs perform better than GPPs. 
It is obvious from above studies that trade-offs are to be expected when a particular design methodology is adopted, hence careful analysis should be carried out beforehand. 
Other comparative studies include \cite{Tian2010,Wu2014,Giefers2016} with similar results and conclusions.

\begin{table*}[]
\centering
\caption{Comparison of SDR Design Approaches}
\label{sdr_comp}
\begin{tabular}{c|c|c|c|}
\cline{2-4}
                                                                 \cline{2-4} 
\multirow{-2}{*}{}                                                        & \textbf{GPP} & \textbf{DSP} & \textbf{FPGA} \\ \hline 
\multicolumn{1}{|c|}{\textbf{Computation}}        & Fixed Arithmetic Engines             & Fixed Arithmetic Engines             & User Configurable Logic               \\ \hline
\multicolumn{1}{|c|}{\textbf{Execution}}          & Sequential                           & Partially Parallel                   & Highly Parallel                       \\ \hline
\multicolumn{1}{|c|}{\textbf{Throughput}}         & Low                                  & Medium                               & High                                  \\ \hline
\multicolumn{1}{|c|}{\textbf{Data Rate}}          & Low                                  & Medium                               & High                                  \\ \hline
\multicolumn{1}{|c|}{\textbf{Data Width}}         & Limited by Bus Width                 & Limited by Bus Width                 & High                                  \\ \hline
\multicolumn{1}{|c|}{\textbf{Programmability}}    & Easy                                 & Easy                                 & Moderate                              \\ \hline
\multicolumn{1}{|c|}{\textbf{Complex Algorithms}} & Easy                                 & Easy                                 & Moderate                              \\ \hline
\multicolumn{1}{|c|}{\textbf{I/O}}                & Dedicated Ports                      & Dedicated Ports                      & User Configurable Ports               \\ \hline
\multicolumn{1}{|c|}{\textbf{Cost}}               & Moderate                                  & Low                                 & Moderate                              \\ \hline
\multicolumn{1}{|c|}{\textbf{Power Efficiency}}   & Low                                  & Moderate                               & High                                  \\ \hline
\multicolumn{1}{|c|}{\textbf{Form Factor}}        & Large                                & Medium                               & Small                                 \\ \hline                                  
\end{tabular}
\end{table*}




\section{Development Tools}
\label{tools}

As we mentioned in Section \ref{fpga_method}, HLS is an abstract method of designing hardware using a high-level programming language.
Developers of FPGA and co-design based SDRs can benefit from HLS as it requires no prior experience with hardware design. 
Unlike the rest of the development tools, HLS tools share a common theme and offer similar features. 
Thus, we first discuss HLS in this section.
Next, we review the common development tools that are typically used in the process of SDR design and implementation for different design approaches.

\subsection{High Level Synthesis (HLS)}
\label{hls}

\begin{figure}[!t]
\centering
\includegraphics[width=0.8\linewidth]{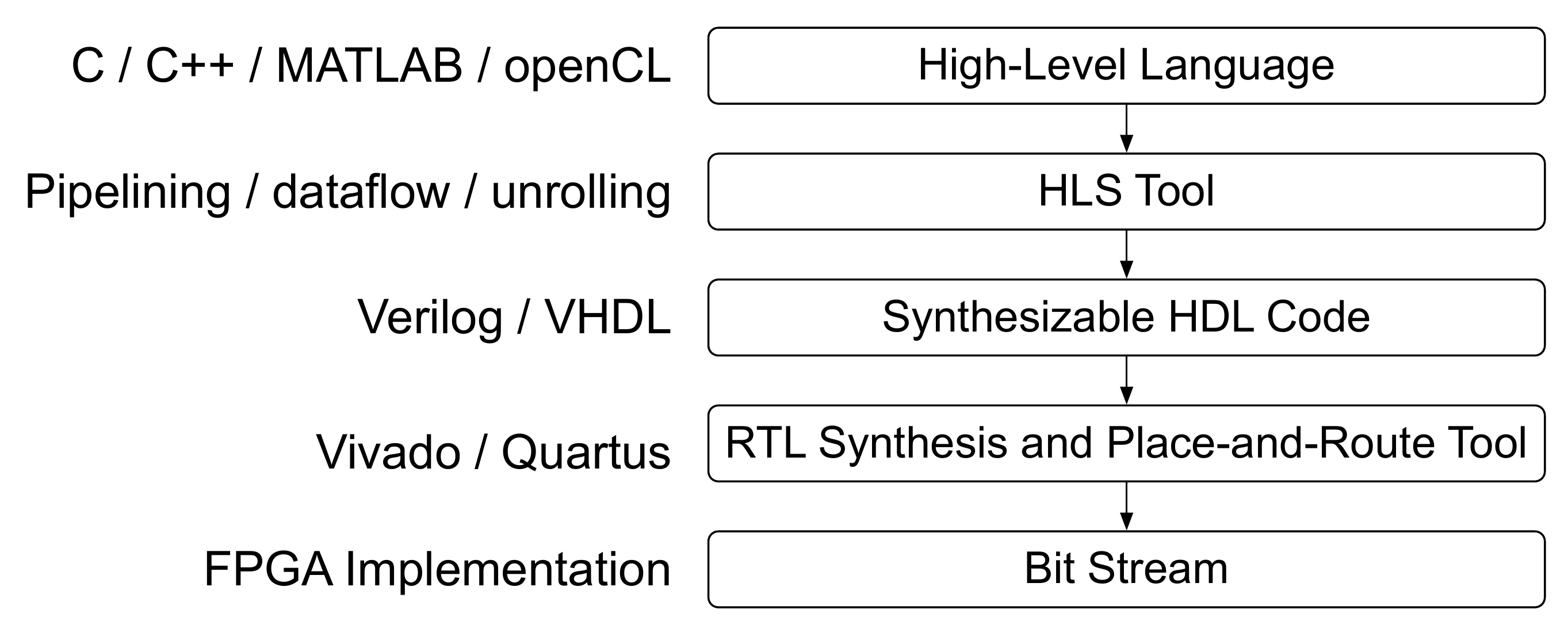}
\caption{HLS design flow.}
\label{hls_fig}
\end{figure}

HLS has been a hot research topic for over a decade, with both academia and industry trying to make hardware design more accessible by every developer \cite{Cardoso2016}. 
HLS is the process of converting an algorithmic specification of the design described by a high-level programming language to an RTL implementation. 
HLS provides a new level of design abstraction through exploring the micro-architecture and any hardware constraints. 
The resulting RTL is highly optimized, in terms of power, throughput and latency, and reasonably comparable to a hand-tuned code. 
Figure \ref{hls_fig} depicts this process. 
The major difference between RTL and C is the absence of timing constraints in the high-level model, which is merely a behavioral description of the system with no details about the underlying hardware. 
The second difference is the processing architecture: while GPP architecture is fixed, the best possible processing architecture is built by the compiler for FPGA \cite{Tessier2015}. 
In addition, HLS can speed up the development cycle (time to market) to several weeks, down from several months \cite{Andrade2017}.
This is because the task of producing an optimized RTL is handled by the HLS tool and the developer's efforts are focused on describing the system's algorithmic description.

In \cite{Canis2013} the authors presented LegUP, an open-source HLS tool. 
This tool is capable of profiling a code to identify frequently executed sections of the code for hardware acceleration (i.e., moving them to the FPGA fabric). 
The authors in \cite{Cong2011} survey HLS compilers and their capability to provide an accurate estimation of functional area and timing, comparable to results from hand-tuned hardware designs.
In an effort to help the developer make the right decision to pick an HLS tool that yields the best results for their application, the authors in \cite{Andrade2017} present a study where they compared three of the industry tools, namely Vivado HLS \cite{hls}, Intel FPGA SDK for OpenCL \cite{intel_sdk}, and MaxCompiler \cite{max_compiler}, through developing LDPC decoders, which are often used as error correcting blocks in SDRs.
All three tools successfully synthesized LDPC decoders and implemented them on Intel \cite{intel} and Xilinx \cite{xilinx} FPGA boards.
The difference, however, was in the logic utilization and performance. 
Similarly, the authors in \cite{Inggs2014}, compare the same aforementioned list of compilers quantitatively and qualitatively using several financial engineering problems (e.g., Monte Carlo-based Option Pricing) and compare the performance of several FPGA boards.
Their results show that both Intel FPGA SDK for OpenCL and MaxCompiler performed better than Vivado HLS due to their ability to extract parallelism more effectively.
In \cite{Nane2016}, the authors comprehensively review recent HLS tools and provide through a careful analysis a methodology based on C benchmarks to compare some of these tools and their optimization features.
The various benchmarks implemented demonstrate that some tools are better suited for certain applications than the rest, with no specific tool dominating the HLS field. 
The authors also show that open-source HLS tools such as LegUP \cite{legup} can be as effective as their commercial counterparts. 
Other surveys and analyses include \cite{Meeus2012},\cite{Windh2015}, \cite{Ravi2016}, which focused on open-source tools, and \cite{Tambara2017}, which studied some of the trade-offs of HLS-generated designs and how reliable they are when errors are injected. 
All of the studies above prove the feasibility and reliability of HLS tools to generate RTL codes, despite having different development and optimization solutions.

Examples of HLS tools include Xilinx Vivado HLS \cite{hls} and SDSoC \cite{sdsoc}, Intel FPGA SDK for OpenCL \cite{intel_sdk}, Cadence Stratus High-level Synthesis (combining Cadence C-to-Silicon and Forte Cynthesizer) \cite{cadence_hls}, Synopsys Synphony C Compiler \cite{synphony}, Maxeler MaxCompiler \cite{max_compiler}, MATLAB HDL Coder \cite{hdl_coder}, and LegUP \cite{legup}, which unlike the rest of the tools is vendor-independent (works with all types of FPGA boards, such as Xilinx \cite{xilinx}, Intel \cite{intel}, Lattice \cite{lattice}, and Microsemi \cite{microsemi}).

Table \ref{hls_tools} presents a summary of commercial HLS tools. 
While some of them are vendor-specific, some tools work with a variety of FPGA boards. 
Examples mentioned in the table all provide a set of area and timing optimizations such as resource sharing, scheduling, and pipelining.
However, not all of them are capable of generating testbenches for the design. 

\begin{table*}[t]
\centering
\scriptsize
\caption{HLS Tools}
\label{hls_tools}
\begin{tabular}{c|c|c|c|c|c|}
\cline{2-6}
                                         & \textbf{\begin{tabular}[c]{@{}c@{}}Xilinx Vivado\\ HLS \end{tabular}}  \cite{hls} & \textbf{\begin{tabular}[c]{@{}c@{}}Intel FPGA \\ SDK for OpenCL \end{tabular}}\cite{intel_sdk} & \textbf{\begin{tabular}[c]{@{}c@{}}Cadence Stratus\\  High-level Synthesis \end{tabular}}\cite{cadence_hls} & \textbf{\begin{tabular}[c]{@{}c@{}}Synopsys Synphony\\  C Compiler \end{tabular}}\cite{synphony} & \textbf{\begin{tabular}[c]{@{}c@{}}Maxeler \\ MaxCompiler \end{tabular}}\cite{max_compiler} \\ \hline
\multicolumn{1}{|c|}{\textbf{Input}}    &          C/C++/SystemC     &     C/C++/SystemC                                                                          &           C/C++/SystemC                                             &          C/C++                                                                        &              MaxJ                                                           \\ \hline
\multicolumn{1}{|c|}{\textbf{Output}}    &       VHDL/Verilog/SystemC                                                                &                                                 VHDL/Verilog                              &         VHDL/Verilog                                                                                 &                      VHDL/Verilog/SystemC                                                           &                           VHDL                                              \\ \hline
\multicolumn{1}{|c|}{\textbf{Testbench}} &          Yes                                                             &     No                                                                          &             Yes                                                                             &               Yes                                                                   &           No                                                              \\ \hline
\multicolumn{1}{|c|}{\textbf{Optimizations}}   &       Yes                                                                &     Yes                                                                         &            Yes                                                                              &                                                    Yes                              &       Yes                                                                  \\ \hline
\multicolumn{1}{|c|}{\textbf{Compatibility}}   &                     Xilinx FPGA                                                  &     Intel FPGA                                                                          &                                                                                    All      &                                                           All                       &                                                           All              \\ \hline
\end{tabular}
\end{table*}

\subsection{Tools}

In this section we review the existing software tools for SDR development. 
For each design methodology, we discuss a compatible development tool and list its features. 
We also provide an overall comparison between them to highlight the differences. 

\begin{table*}[]
\centering
\scriptsize
\caption{Development Tools and Platforms}
\label{dev_tools}
\begin{tabular}{c|c|c|c|c|c|c|}
\cline{2-7}
                                                                & \textbf{MATLAB \& Simulink} \cite{mathworks}& \textbf{Vivado HLS \& SDSoC} \cite{xilinx}& \textbf{LegUP} \cite{legup}& \textbf{GNU Radio} \cite{gnu}& \textbf{LabView} \cite{labview}& \textbf{CUDA} \cite{cuda}\\ \hline
\multicolumn{1}{|c|}{\textbf{Input}}    & MATLAB/Graphical                               & C/C++/SystemC        &     C                          & Graphical/Python/C++                       & Graphical                                & C/C++/Fortran/Python                         \\ \hline
\multicolumn{1}{|c|}{\textbf{Output}}   & MATLAB/C++/RTL                                      & C/RTL           &           C/RTL                         & C/RTL                                      & C/RTL                                    & Machine Code                          \\ \hline
\multicolumn{1}{|c|}{\textbf{Platform}} & GPP/GPU/DSP/FPGA                                    & GPP/FPGA             &         GPP/FPGA                       & GPP/GPU/DSP/FPGA                           & GPP/GPU/DSP/FPGA                         & GPU                                   \\ \hline
\multicolumn{1}{|c|}{\textbf{Licence}}  & commercial                                          & commercial          &         open-source                        & open-source                                & commercial                               & commercial                            \\ \hline
\end{tabular}
\end{table*}

\subsubsection{MATLAB and Simulink}

\begin{figure}[!t]
\centering
\includegraphics[width=0.9\linewidth]{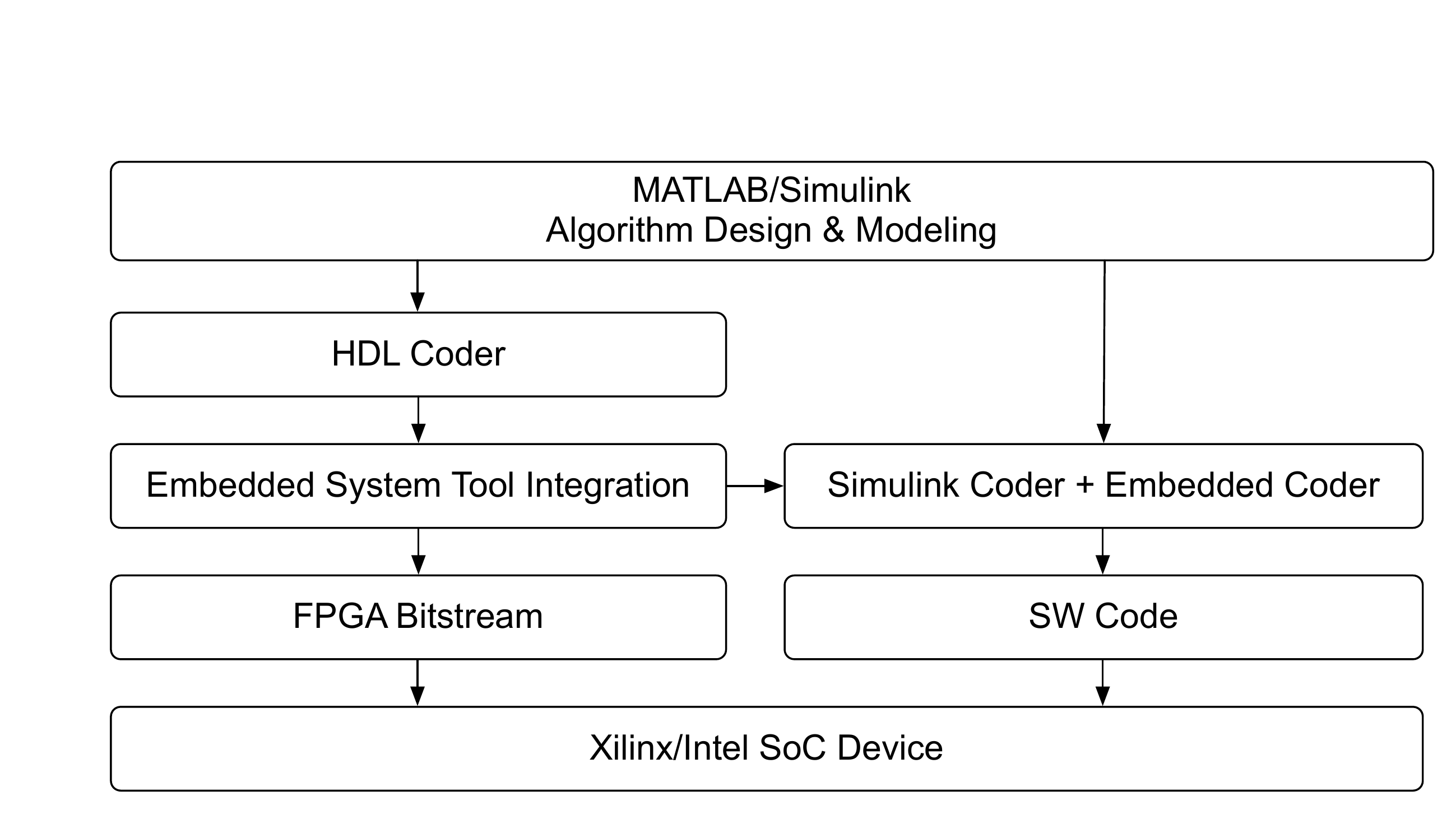}
\caption{Mathworks SoC design flow \cite{mathworks}.}
\label{matlab_soc}
\end{figure}

Most designers start with modelling and simulating the system using Mathworks MATLAB \cite{matlab} and Simulink \cite{simulink}. Through the availability of a wide range of built-in functions and toolboxes, especially for signal processing and communication, developing and testing applications became very common and widely adopted. However, in order to use these models for different platforms, developers would need to use MATLAB Coder \cite{coder} and Simulink Coder \cite{sim_coder} to generate C/C++ codes. The generated codes can be used with Embedded Coder \cite{embed_coder} to optimize them and generate software interfaces with AXI drivers for the sake of running on embedded processors and microprocessors, like the dual ARM cortex A9 MPcore \cite{arm} on the ZedBoard \cite{zedboard}. Alternatively, developers can use the HDL Coder \cite{hdl_coder} to generate synthesizable RTL (Verilog or VHDL) code to be implemented on FPGAs or ASICs.  It also has support for Xilinx \cite{xilinx} and Intel \cite{intel} SoC devices by providing some information and optimizations pertaining to resource utilization and distributed pipelining. Figure \ref{matlab_soc} shows the design flow for SoC platforms that the aforementioned tools offer and how they are connected. Examples of using MATLAB and Simulink to develop an SDR are the works by \cite{Stewart2015a} and \cite{Stewart2015b}, where the authors used the RTL-SDR very low-cost SDR dongle \cite{rtl-sdr} ($\sim\$20$) with a desktop computer to design an academic curriculum for teaching DSP and communications theory. 

\subsubsection{Vivado HLS and SDSoC}

Xilinx Vivado HLS \cite{hls} is a design environment for high-level synthesis. 
This tool offers a variety of features to tweak and improve the RTL netlist output that is compatible and optimized for Xilinx FPGA boards. 
It accepts input specifications described in several languages (e.g., C, C++, SystemC, and OpenCL), and generates hardware modules in Verilog or VHDL. 
Developers have several options to optimize the solution in terms of area and timing through the use of directives (guidelines for the optimization process) and pragmas for RTL optimization. 
These optimizations include loop unrolling, loop pipelining, and operation chaining. 
SDSoC \cite{sdsoc} is another tool by Xilinx \cite{xilinx}. 
The major difference between the two tools is that the latter has the capability to provide solutions for SoCs. 
SDSoC is built on top of Vivado HLS and has the same C-to-RTL conversion capability. 
The main advantage of using SDSoC is that it automatically generates data movers, which are responsible for transferring data between the software on the processor and the hardware on the FPGA.

A similar tool to SDSoC that is open-source is LegUP \cite{legup}. It was developed at the University of Toronto, as part of an academic research effort to design an HLS tool that is capable of taking in a C code as an input and providing three possible outputs: a synthesizable RTL code for an FPGA, a pure software executable, and a hardware/software co-design solution for a SoC.

\subsubsection{GNU Radio}

It is an open-source software development toolkit that provides signal processing blocks to implement SDRs \cite{gnu2004,gnu}. It runs on desktop or laptop computers, and with the addition of simple hardware such as USRP B200 \cite{usrp}, can build a basic SDR. 
It is often used by academia and the research community for simulation as well as quick setup of SDR platforms. 
Similar to System Generator tool \cite{sys_gen} and Simulink \cite{simulink}, it includes different kinds of blocks such as decoders, demodulators, and filters. 
It is also capable of connecting these blocks and managing data transfer in a reliable fashion.
In addition, it supports USRP systems \cite{usrp}. 
One of the attractive features of GNU Radio is the ability to define and add new blocks. 
This can be done via programming in C++ or Python. 
An example of using GNU Radio is the work by \cite{Malsbury2013}, where the author uses it with a USRP to realize different types of transceivers such as TDMA and CSMA, and showcases some of its capabilities. 
Similarly, the authors in \cite{Abirami2013} successfully achieve real-time communication between two computers using USRP \cite{usrp} and RTL-SDR \cite{rtl-sdr}.

\subsubsection{LabVIEW}
 
A widely used tool from National Instruments \cite{labview} that offers a visual programming environment for test, automation and control applications used by both industry and academia. It is similar to GNU Radio and Simulink, where the design can be realized schematically by connecting a chain of various blocks together, each of which performs a certain function. It also offers complete support for USRP \cite{usrp} enabling rapid prototyping for communications systems. 
Designing different blocks of the system can be achieved using high-level languages such as C or MATLAB, or using a graphical dataflow. 
An SDR platform development using LabVIEW is the work by \cite{Kucuk2017}, where the author describes a wireless communication course design that incorporates USRP and LabVIEW, due to their ease of use, to help teach students basic concepts.
Similarly, in \cite{Jimenez2017} the authors designed an SDR platform, namely FRAMED-SOFT, that includes two types of USRPs and is intended for an academic environment.

\subsubsection{CUDA}

Developed by NVIDIA, it issues and manages computing platforms and programming models for data-parallel computing on GPUs \cite{cuda}. 
Developers typically use CUDA when GPUs are part of the processing architecture as co-processors, and want to take full advantage of their power by speeding up applications. 
As discussed in Section \ref{gpu}, in order to identify application components that should run be on GPP and the parts that should be accelerated by the GPU, one needs to look at the tasks at hand. 
Programming languages that can be used in CUDA include C, C++, Python, Fortran, and MATLAB \cite{matlab}. 
The toolkit includes, in addition to the rich library for GPU-related acceleration functions, a compiler, development tools, and CUDA runtime, to develop applications and optimize them for systems that incorporate GPUs.

\section{Platforms}
\label{plat}

In this section, we list the different types of SDRs from the architecture and design point of view. We analyze them, examine their strengths and shortcomings, and discuss their impact on SDR development. 

\subsection{GPP-based} 

\textbf{USRP N-Series.} Universal Software Radio Peripheral (USRP) is the most common SDR platform known to the developers' community \cite{usrp}. 
The cost of this platform is around \$4000-5000.
It provides a hardware platform for the GNU Radio Project \cite{gnu2004}. 
There are two generations available: USRP1 and USRP2.
USRP1 (released in 2004) is connected to a generic computer through USB, with the addition of a small FPGA. 
The FPGA board has two roles: routing information, and limited signal processing. 
This generation was capable of supporting a $\sim3$MHz bandwidth due to USB 2.0 limitation. 
The second generation, USRP2, was released in 2008, and it supports 25MHz bandwidth by utilizing gigabit Ethernet. 
It includes a Xilinx Spartan 3 FPGA \cite{xilinx} for local processing operations.

\begin{figure}[!t]
\centering
\includegraphics[height=1.8in,width=2.8in]{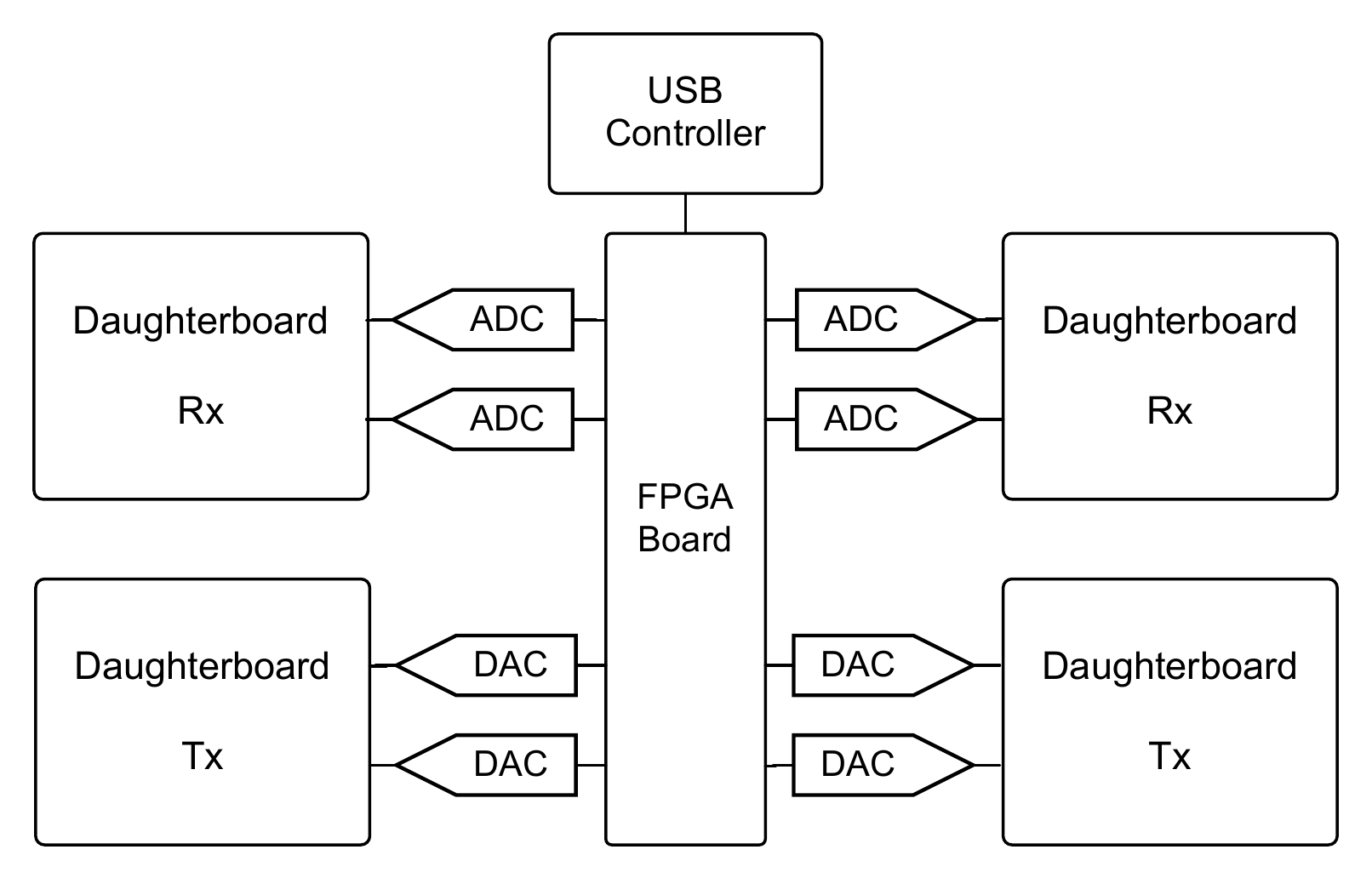}
\caption{USRP board architecture. RF daughterboard selection depends on the application specifications in terms of frequency coverage.}
\label{usrp}
\end{figure}

USRP, in general, is a board with ADC and DAC, an RF front end, a PC host interface, and an FPGA. 
This board consists of a motherboard and typically four daughterboards (two transmitters Tx, and two receivers Rx), as depicted in Figure \ref{usrp}. 
The daughterboards process analog operations like filtering and up/down conversions. 
They are modular so they can deal with applications operating up to 6GHz. 
The FPGA board, depending on the USRP series, handles a few signal processing operations, and the majority of operations are offloaded to the connected host system.
USRP platforms can be easily set up to use. 
However, while their performance is suitable for research experiments and quick prototyping, these platforms do not necessarily meet the requirements of communication standards.
In fact, the minimum bandwidth of the RF, PC host, or FPGA component used affects the throughput and timing characteristics of the platform.

\textbf{KUAR.} Another platform that is similar to USRP is the Kansas University Agile Radio (KUAR) platform \cite{kuar2007}. 
The basic architecture is composed of a generic computer and a Xilinx Virtex II Pro P30 FPGA, which has two PowerPC 405 cores \cite{xilinx}. 
The main advantage of this platform is the degree of flexibility that it offers. 
Developers have the option to implement communication standards in three different methods: (i) they can fit the entire design onto the FPGA and assign few tasks to the host's GPP (full hardware), (ii) a full software implementation, where the design is implemented entirely on the GPP with minimal FPGA involvement, and (iii) a hybrid hardware/software co-design implementation, where the developer can partition the design in any way that fits their criteria.

\textbf{LimeSDR.} Resembling the general basic architecture of USRP (e.g., USRP B200 \cite{usrp}), Lime Microsystems \cite{lime} develops a series of SDRs that is based on Lime Microsystem’s latest generation of field programmable RF transceiver (FPRF) technology, in addition to an Intel FPGA \cite{intel} and a microcontroller. 
It is then connected to a computer via USB 3.0. 
The SDR platform has the responsibility of delivering the wireless data, while the GPP (computer) has the task of processing the incoming signals and generating the data to be transmitted by the SDR. 
The GPP in this case is the source of computing power for baseband. 
LimeSDR supports signal bandwidth $\sim$30-60MHz. 
Developers of LimeSDR also developed LimeSuite software, which is used to model SDRs. This tool, source code, firmware, and schematics, are open-source.

\textbf{Ziria.} It is a programming platform that uses a domain-specific language (DSL) named Ziria, and an optimizing compiler \cite{Stewart2015}. 
The application of Ziria is the implementation of the physical layer (PHY) of wireless protocols. 
Ziria has a 2-layer design:  
\begin{itemize}
\renewcommand\labelitemi{--}
\item A lower layer that is an imperative language, which is a mixture of C and MATLAB \cite{matlab} code, with features of the two languages carefully chosen to guarantee efficient compilation.  
\item A higher layer, which is the language used to specify and stage stream processors. 
\end{itemize}

The Ziria optimizing compiler consists of two parts: the frontend and the backend. 
The frontend parses the Ziria code, putting it in an abstract representation, then applying several optimizations on it. The backend compiles the resulting optimized code into an optimized low-level execution model. 

The particular benefits of this platform are as follows:
The first one is easy and dynamic reconfiguration, due to the dynamic staging of the control graph.
This is unlike GNU Radio \cite{gnu} C++ templates that allow limited reconfigurability. 
In addition, its code optimization can operate on data-flow components, and often yields a faster execution on GPPs (e.g., through the use of LUTs). 
In general, a Ziria code is very concise and easy to use. 
For example, an implementation of a WiFi scrambler in Ziria needs thirteen lines only. 
Ziria is an interesting research based on data-flow languages, which are typically used in embedded systems. 
It also builds upon popular functional reactive programming framework.

\textbf{Sora.} Sora is a fully programmable software radio platform on PC architecture. 
It requires C programming on multi-core GPP, and yields high performance that includes high processing speed and low latency. 
The Sora platform uses the Ziria language discussed above to write high-level SDR descriptions, and is tested for real-time operation \cite{Tan2011}. 
Unlike WARP \cite{warp} (which we will explain in Section \ref{ex_hybrid}), Sora can accommodate various RF front ends.

Since PC hardware is not intended for signal processing of wireless protocols, their performance can be limited. 
We discussed some of the limitations of GPPs in the context of SDRs in Section \ref{gpp_sec}.
These limitations were the motivation behind the development of Sora.
The overall setup of Sora includes a soft-radio stack that combines a multi-core GPP and a radio control board, which consists of a Xilinx Virtex-5 FPGA \cite{xilinx}, PCIe-x8 interface, and DDR2 SDRAM. 
Sora uses both hardware and software techniques to address the challenges of using PC architectures for high speed SDR. It is the first SDR platform that enables users to develop high speed wireless implementations entirely in software on a standard PC architecture.

In Sora, new techniques are proposed  for efficient PHY implementation. 
Some of these techniques include: (i) exploiting large high-speed cache memory to minimize memory accesses, (ii) extensive use of LUTs, where they would trade memory for calculation and still well fit into L2 cache, (iii) exploiting data parallelism in PHY, and (iv) utilizing wide-vector SIMD extension in GPP. 
One of the main novelties of Sora is efficiently partitioning and scheduling the PHY processing across cores in GPP. 
The second innovation is core dedication for real-time support. 
This was accomplished by exclusively allocating enough cores for SDR processing in multi-core systems. 
They showcased its effectiveness through SoftWiFi, which is a full implementation of IEEE 802.11a/b/g PHY and CSMA MAC with 9000 lines of C code and real-time performance. 
They also successfully implemented a 3GPP LTE Physical Uplink Shared Channel (PUSCH), or Soft-LTE, with 5000 lines of C code and a peak rate of 43.8Mbps with 20MHz.

%
 
Some of the critiques to Sora is that its FPGA is not programmable and its capabilities are not fully utilized. 
In addition, the authors do not share the details of its internal routines. 
Finally, Sora only works on GPPs, with no clear mapping to DSPs.

\textbf{Iris.} It is a cross-platform SDR, developed to support highly reconfigurable radio networks \cite{Sutton2010}. 
It is built using a plugin architecture, namely \textit{components}, to achieve modularity. 
These components process data streams and perform different operations on them. 
An engine is used to run, load and maintain the components.
Similar to System Generator \cite{sys_gen} and Simulink \cite{simulink} tools, Iris engine can be used to link components together to build a complete radio system. 
XML is used to specify the components used in the program, their parameters, and how they should be linked.  
The main features of the Iris architecture include: (i) runtime configurability, (ii) support for the entire network stack (all layers), not just the PHY layer, (iii) support for advanced processing platforms including FPGAs.

There are multiple studies on using Iris.
An example is the work in \cite{belt2013}, where the authors discuss the process through which they were able to implement Iris on Xilinx Zynq SoC \cite{xilinx}. 
The motivation behind this work is the fact that communication systems are in a constant state of development and they increase in complexity and sophistication. 
This calls for higher computational performance and, along with it, a higher level of flexibility. 
In order to realize Iris on Zynq platform, the components are first translated into C++ using Cmake tools and then they are ported to the platform. 
HLS tools, like Xilinx HLS \cite{hls}, can be used to accelerate parts of the design that are considered to be the bottlenecks. 
This is done by running system profiling on the software components, and one of these profilers is Linux Perf. 
Acceleration, in particular, refers to running parts of the software (after re-writing it in RTL) on the FPGA fabric in order to achieve higher performance.

\subsection{GPU-based}


\textbf{\\WiMAX SDR.} The authors in \cite{gpu2010} built a GPU-based platform to realize a WiMAX system. 
In their study, they also compared the performance of GeForce 9800GTX GPU \cite{nvidia} against a TMS320C6416 DSP \cite{ti} via implementing the Viterbi decoder algorithm.
The results indicate that the throughput of the GPU is 181.6Mbps, with a considerable difference compared to the DSP's 2.07Mbps. 

\begin{figure}[!t]
\centering
\includegraphics[width=0.75\linewidth]{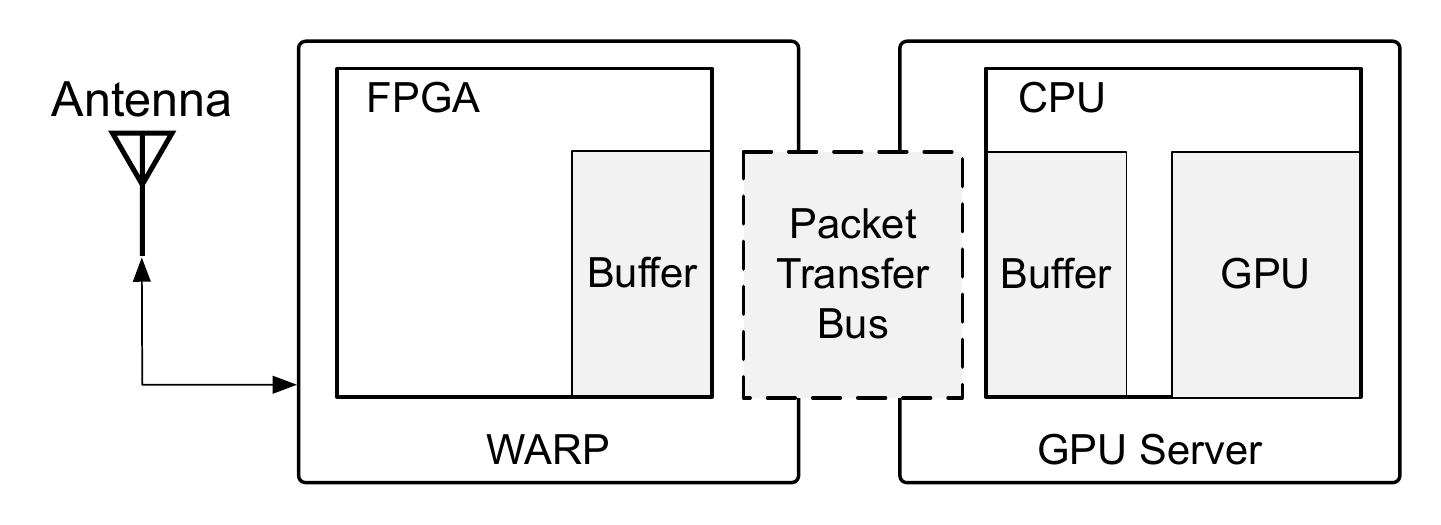}
\caption{GPU-accelerated SDR Platform using WARP \cite{Li2014}. The GPU server is used for baseband processing, while WARP is used for radio control and interface. Offloading signal processing tasks to the GPU has significantly improved the overall performance.}
\label{gpu_sdr}
\end{figure}

\textbf{\\OFDM for WiFi Uplink SDR.} In \cite{Li2014}, the authors used the WARP framework \cite{warp} as a basis to realize their NVIDIA GPU-based SDR platform. 
They utilized the inherent parallelism of GPUs, and with the help of CUDA \cite{cuda}, they were able to achieve real-time performance on WARP. 
They used this platform to design and implement a Single Input Single Output (SISO) OFDM system for WiFi uplink communication. 
Figure \ref{gpu_sdr} depicts the architecture of this enhanced (accelerated) WARP SDR. 
For this platform, they used: (i) a WARP version 3, which consists of a Xilinx Virtex-6 FPGA \cite{xilinx} for radio control and interface, and (ii) a GPU server, which consists of an Intel i7-3930K six-core 3.2GHz CPU \cite{intel} for transceiver configuration, and four NVIDIA GTX TITAN graphic cards \cite{nvidia} for baseband processing. 
Each TITAN is comprised of 2880 core Kepler GPU running at 889MHz. 
The accelerated WARP achieves less than 3ms latency and higher than 50Mbps over-the-air throughput.

\textbf{\\Signal Detection SDR.} 
In \cite{Fisne2017} the authors designed and studied real-time signal detection using an SDR platform composed of a laptop computer with an Intel Xeon E3-1535M processor \cite{intel} and an NVIDIA Quadro M4000M GPU \cite{nvidia}. 
For 1000ms long samples, this design achieves around 75$\%$ reduction in parallel processing time, compared to GPPs.

\subsection{DSP-based}

\textbf{Imagine Processor-based SDR.} 
Authors of \cite{Rajagopal} proposed one of the earliest SDR solutions that is fully based on a DSP. 
This SDR employs the Imagine stream processor, developed at Stanford University in 2001 \cite{khailany2001}. 
The Stanford Imagine project aimed at providing a signal and image processor that was C programmable and was able to match the high performance and density of an ASIC. 
It is based on stream processing \cite{khailany2003,gumma2005,khailany2008}, which is similar to dataflow programming in exploiting data parallelism and is suitable for signal processing applications. 
This work paved the way to the development of GPUs.

\begin{figure}[!t]
\centering
\includegraphics[height=1.9in,width=3.2in]{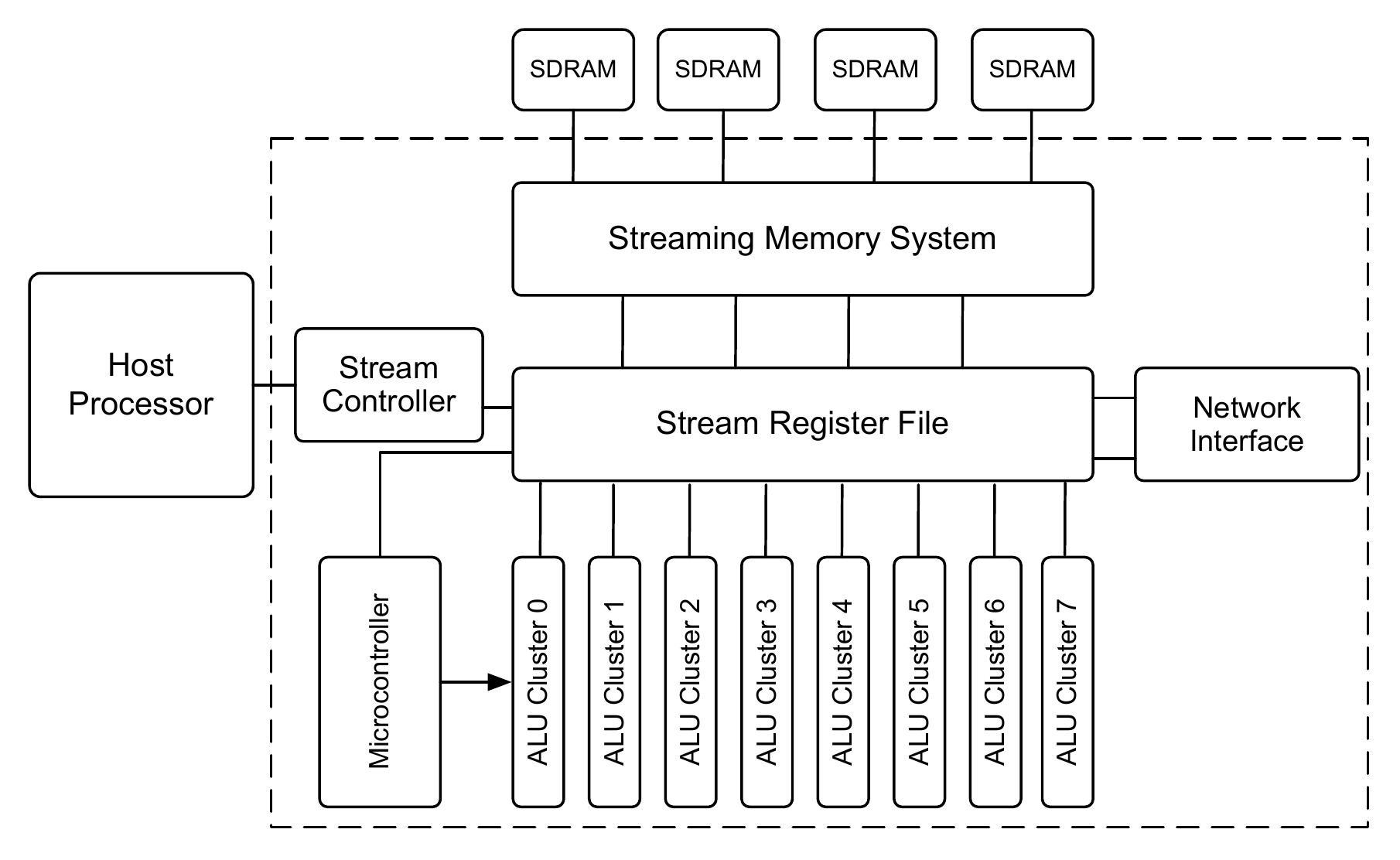}
\caption{Imagine Processor Architecture. This platform employs a real-time stream processor for baseband processing.}
\label{imagine}
\end{figure} 

As Figure \ref{imagine} shows, Imagine processor uses VLIW-based ALU clusters arranged in a SIMD fashion to handle data streams. 
At the middle of the architecture, there is the Stream Register File, which stores data from other components, thereby minimizing memory accesses. 
The performance of this platform has been evaluated through implementing complex algorithms relevant to W-CDMA cellular system. 
The results show higher performance compared to TI C67 DSP \cite{ti}: channel estimation and detection are improved by 48x and 42x, respectively.

\textbf{SODA.} In \cite{Lin2006}, the authors introduce the Signal-processing On-Demand Architecture (SODA), which is an SDR platform based on multi-core DSPs. 
It offers full programmability and targets various radio standards. 
The SODA design achieves high performance, energy efficiency, and programmability.
This is attributed to a combination of features that include SIMD parallelism and hardware support for 16bit computations, since most algorithms operate on small values. 
The basic processing element is an asymmetric processor consisting of a scalar and SIMD pipeline, and a set of distributed scratchpad memories that are fully managed in software. 
SODA is a multi-core architecture, with one ARM Cortex-M3 processor \cite{arm} for control purposes and multiple processing elements for DSP operations.
Using four processing elements can meet the computational requirements of 802.11a and W-CDMA. 
Compared to WARP and Sora, as a single-chip implementation, SODA is more appropriate for embedded scenarios. 
As with WARP, developers must learn the architecture in order to implement SDRs.

\textbf{ARM Ardbeg.} In \cite{Woh2008}, a commercial prototype based on SODA architecture was presented. 
The main enhancements of Ardbeg compared to SODA are optimized SIMD design, VLIW support, and a few special ASIC accelerators, which are dedicated to certain algorithms such as Turbo encoder/decoder, block floating point and arithmetic operations.

\textbf{Atomix.} Typically, programmers need to do one of three tasks to the software workflow of DSPs: tapping into a signal processing chain, tweaking a block, or inserting/deleting a block. 
To simplify these tasks, modularity is crucial.
However, designing a modular software for DSPs is challenging considering the particular requirements, such as latency sensitivity and high throughput, that must be supported. 
The main challenge is the need for programmers to define and manage everything manually and explicitly. 
In other words, it is necessary to use bare metal features, like moving data across cores, managing SRAMs, and parallelizing software.  

In order to address these concers, Atomix \cite{Bansal2015} describes the software in blocks, named atoms. 
An atom can be used to implement any operation.
This can be signal processing or system handling. 
Atoms can be used for realizing blocks, flowgraphs, and states in wireless stacks. In addition, simple control flow makes atoms composable. 
The easy modification feature mentioned above is due to declarative language. 
It is important to note that an Atomix signal processing block implements a fixed algorithmic function, operates on fixed data lengths, is associated with a specific processor type, and uses only the memory buffers passed to it during invocation. 
The blocks will run fixed sets of instructions executing uninterrupted on fixed resources using fixed memories. 
This results in having fixed execution times. 
Atoms can also be composed to build larger atoms. 
Using Atomix, radio software can be built entirely out of atoms and is easily modifiable. 
Atomix-based radio also meets throughput and latency requirements. 

Developers define the atoms in C.
Then, the blocks are composed into flowgraphs and states using Atomix interface. 
The next step involves developing a parallelized schedule and resource assignment in order to meet latency and throughput requirements. 
The software is then compiled by Atomix into a low-level C. 
The compiled code, along with Atomix libraries, will be compiled into a binary. 
Atomix was used to build IEEE 802.11a receiver only, namely Atomix11a.
Evaluation of Atomix11a shows that it exceeds receiver sensitivity requirements, operates in indoor environments robustly, has low processing latency, and atoms have low timing variability. 
Although the power consumption reported is 7W, it does not include the power consumed by the front end, USRP2, which is about 14W. 
A shortcoming of Atomix is that it is intended only for synthesis on a variety of DSPs, not for GPPs, GPUs, or FPGAs.

\textbf{BeagleBoard-X15.} 
A collaborative project between Texas Instruments \cite{ti}, Digi-Key \cite{digi}, and Newark element14 \cite{newark}, BeagleBoard is an open-source SoC computer \cite{beagle}. 
It features TI Sitara AM5728 \cite{ti}, which includes two C66x DSPs \cite{ti}, two ARM Cortex-A15, two ARM M4 cores \cite{arm}, and two PowerVR SGX544 GPUs \cite{imagination}. 
With its relatively low price ($\sim$\$270), the DSPs along with the co-processors make a powerful platform for implementing standalone SDRs. 
An example of using the BeagleBoard (an older model but same general architecture) is the work by \cite{Fayez2011}, where it was used to implement the Public Safety Cognitive Radio (PSCR) \cite{pscr} through GNU Radio \cite{gnu}.
PSCR is FM radio-based and was developed by the Center for Wireless Telecommunications (CWT) at Virginia Tech.

\subsection{FPGA-based}

\textbf{Airblue.} 
This work \cite{Ng2010} introduces an FPGA-based SDR platform for PHY and MAC layers. 
Airblue is a method to implement radios on FPGA to achieve configurability.
This is done using an HDL language called Bluespec, through which all hardware blocks of a radio transceiver are written. 
In Bluespec, a developer describes the execution semantics of the design through Term Rewriting Systems (TRS). 
TRS is a computational paradigm based on the repeated application of simplification rules \cite{Bezem2003}. 
The next step is compiling the TRS into RTL codes. 
TRS has the capability of generating efficient hardware designs. 
The main difference between a Verilog interface and a Bluespec interface is that the former is merely a collection of wires with no semantic meaning, while the latter includes handshake signals for blocks communication. 
Therefore, Bluespec facilitates latency insensitive designs which are essential to system construction via modular composition. 
Using Airblue, developers may find the need to modify the building blocks, or modules, to add new features, make algorithmic modifications, tune the performance to meet throughput or timing requirements, or make FPGA-specific optimizations.  
 
In order to reach modular refinements, the design of a configurable radio must have two design properties, latency-insensitivity and data-driven control. 
In addition to flexibility, Airblue meets tight latency requirements by implementing both PHY and the MAC on FPGA and connecting them with streaming interfaces, which allows data to be processed as soon as it is ready. 
Another advantage of Airblue is the implementation of highly reusable designs through parameterizations (i.e., same RTL block can be instantiated several times using different parameter values). 
Also, several techniques that permit designers to reuse existing designs to implement run-time parameterized designs are developed. 
 
Airblue essentially is a HLS platform that offers modular refinement, where modifying one module does not affect the rest of the system, similar to the approach adopted by Atomix.
This is why, when compared to WARP, Airblue is more flexible, since WARP was not designed with the aforementioned principles in mind. 
The authors in \cite{Ng2010} also studied HLS tools and compared them to Bluespec.
They found them to be more effective than Bluespec in early stages of the design; nevertheless, Bluespec is capable of yielding a more optimized final RTL. 
They argued that HLS will be of limited use in final FPGA implementations, especially for high performance blocks required by future wireless protocols, hence Bluespec is the language of choice for Airblue. 
For performance evaluation, the authors have implemented 802.11 and experimented with a set of protocol changes. 
Airblue demonstrated that it was easily modifiable and still meets timing requirements.
Airblue achieves a higher speed than Sora for cross-layer communication (between MAC and PHY layers), which typically has the latency requirement of a few microseconds.
 
%


\subsection{Hybrid Design}
\label{ex_hybrid}

\textbf{USRP Embedded (E) Series.} 
This is the embedded version of USRP, where they incorporate Xilinx SoCs \cite{xilinx} to develop standalone SDRs. 
USRP E310, for example, is based on Xilinx Zynq 7020, which yields high performance and is energy efficient. 
This USRP is stand-alone and suitable for mobile applications. 
It supports frequency range 70MHz to 6GHz and features a $2\times2$ MIMO RF front end.

\textbf{WARP.} The Wireless open-Access Research Platform is another example of a co-design specifically developed to prototype wireless protocols \cite{warp}. 
It is programmable and scalable, however, parts of the device are implemented in ASIC, which makes it less flexible. 
WARP v3 contains a Xilinx Virtex-6 FPGA \cite{xilinx}, which includes two MicroBlaze processors and a Gigabit Ethernet peripheral. 
It requires the use of Xilinx Embedded Development Kit (EDK) \cite{xilinx} to design SDRs. 
It is also open-source, with 802.11 reference design made available to the research community. 
WARP has become widely used in the research community due to its effectiveness in implementing various wireless protocols (e.g., 802.11 g/n PHY and MAC) and achieving real-time performance. 
It also supports MIMO since it has multiple RF interfaces.

%

\textbf{PSoC 5LP.} Developed by Cypress Semiconductor \cite{psoc}, this SoC is composed of an ARM Cortex-M3 GPP \cite{arm}, an analog system, and a digital system that uses Universal Digital Blocks (UDBs). 
A UDB is a programmable digital block based on  Programmable Logic Devices (PLDs) for realizing synchronous state machines, i.e., they resemble an FPGA, but are smaller. 
All parts are reconfigurable and programmable by using the PSoC Creator software IDE, which includes a graphical design editor. 
A few developers have used it to build standalone SDRs with simplicity \cite{psoc2}. 
In addition to its low price ($\$15$), PSoC is a good candidate to quickly prototype and get familiar with SDRs.

\textbf{Zynq-based SDR.} The authors in \cite{Drozdenko2017} developed an SDR using the Xilinx Zynq ZC706 and ZedBoard SoCs to implement IEEE 802.11a. 
For their RF front end, they used Analog Devices FMComm3 AD9361 \cite{AD}. 
They used two Zynq boards to compare their performances. 
They both include dual-core ARM Cortex-A9 \cite{arm}, with ZC706 containing Kintex-7 and ZedBoard containing Artix-7 FPGAs. 
In addition to HDL Coder \cite{hdl_coder} and Embedded Coder \cite{embed_coder} to generate RTL and software executable, they used Mathworks Simulink \cite{simulink} to generate the model. 
They reported an average of 2W power consumption for Tx and Rx, compared to 5 to 8W reported by Atomix \cite{Bansal2015}.

\subsection{Comparison}

Table \ref{comp_sdr} compares the SDR platforms discussed above in an effort to provide a reference guide for developers. SDR platforms are compared according to the following criteria:
\begin{itemize}
\renewcommand\labelitemi{--}
\item \textit{Programmablity}: As protocols evolve, a platform is re-programmed by simply adding or exchanging parts of the design. 
For example, when 3GPP issues an update for LTE standard, instead of replacing the entire radio, an SDR is reprogrammed to accommodate the upgrade. 
\item \textit{Flexibility}: A platform is capable of handling future wireless protocols as requirements become more demanding. This means an SDR should be able to support tighter timing requirements for next generation protocols. 
\item \textit{Portability}: A platform is standalone and readily deployed. This is generally a requirement for mobile and IoT applications. 
\item \textit{Modularity}: A design's components are separated and recombined without internal module changes. 
For example, a developer may need to exchange a Viterbi decoder with a Turbo decoder without having to worry about the rest of the design. 
\item \textit{Computing Power}: Since performance depends on the protocol used, we merely evaluate the capability of platforms to implement a given protocol, and not be limited to a subset of the protocol (e.g., implementing the Viterbi decoder only).
\item \textit{Energy Efficiency}: Similar to computing power, we evaluate the capability to implement protocols efficiently, while keeping power consumption at a minimum.
\item \textit{Cost}: The cost of the hardware equipment. 
If a platform requires a PC, its cost is not included. 
Also, when the price of the entire setup is unknown, the price of the basic platform (not including RF front end or interfaces) is shown. 
\end{itemize}
\begin{table*}[!ht]
\centering
\scriptsize
\caption{Comparison of Existing SDR Platforms}
\label{comp_sdr}
\renewcommand{\arraystretch}{1.1}
\setlength{\tabcolsep}{0.7em} 
\begin{tabular}{c|c|c|c|c|c|c|c|c|c|}
\cline{2-10}
                                             & \begin{tabular}[c]{@{}c@{}}\textbf{Programm-} \\ \textbf{ability}\end{tabular} & \begin{tabular}[c]{@{}c@{}}\textbf{Flex-} \\ \textbf{ability}\end{tabular} & \begin{tabular}[c]{@{}c@{}}\textbf{Port-} \\ \textbf{ability}\end{tabular} & \begin{tabular}[c]{@{}c@{}}\textbf{Modul-} \\ \textbf{arity}\end{tabular} & \begin{tabular}[c]{@{}c@{}}\textbf{Computing} \\ \textbf{Power}\end{tabular} & \begin{tabular}[c]{@{}c@{}}\textbf{Energy} \\ \textbf{Efficiency}\end{tabular}  & \textbf{Soft Core} & \textbf{FPGA} & \begin{tabular}[c]{@{}c@{}}\textbf{Cost} \\ \textbf{(USD)}\end{tabular} \\ \hline
\multicolumn{1}{|c|}{\textbf{Imagine-based}\cite{Rajagopal}}  & $\checkmark$  &  $\times$&    $\times$&     $\times$ &         Medium                 &  Low & \begin{tabular}[c]{@{}c@{}}Imagine Stream \\ Processor\end{tabular}   &   N/A    &  N/A   \\ \hline
\multicolumn{1}{|c|}{\textbf{USRP X300}\cite{usrp}}          &   $\checkmark$ &  $\checkmark$ &     $\times$                &       $\checkmark$              &   High &  Low &  PC &  \begin{tabular}[c]{@{}c@{}}Xilinx \\ Kintex-7\end{tabular}  &  \begin{tabular}[c]{@{}c@{}}$\sim4-5K$ \\ Total\end{tabular}  \\ \hline
\multicolumn{1}{|c|}{\textbf{USRP E310}\cite{usrp}}          &   $\checkmark$ &   $\checkmark$ &     $\checkmark$        &       $\checkmark$              &       High                   &                           High &    \begin{tabular}[c]{@{}c@{}}Dual-core ARM \\ Cortex-A9\end{tabular}          & \begin{tabular}[c]{@{}c@{}}Xilinx \\ Artix-4\end{tabular}   &  \begin{tabular}[c]{@{}c@{}}$\sim3K$ \\ Total\end{tabular} \\ \hline
\multicolumn{1}{|c|}{\textbf{KUAR}\cite{kuar2007}}          &$\checkmark$ & $\times$&                $\times$& $\times$&   Medium   &   Low    &  \begin{tabular}[c]{@{}c@{}}Pc + \\ $2\times$ PowerPC cores\end{tabular}            & \begin{tabular}[c]{@{}c@{}}Xilinx \\ Virtex II Pro\end{tabular}  & N/A  \\ \hline
\multicolumn{1}{|c|}{\textbf{LimeSDR}\cite{lime}}          &   $\checkmark$ &  $\checkmark$ &     $\times$                &       $\checkmark$              &   High &  Low & \begin{tabular}[c]{@{}c@{}}PC \\ \end{tabular}  &  \begin{tabular}[c]{@{}c@{}}Intel \\ Cyclone IV\end{tabular}  &  \begin{tabular}[c]{@{}c@{}}$\sim300$ \\ Board Only\end{tabular}  \\ \hline
\multicolumn{1}{|c|}{\textbf{Ziria}\cite{Stewart2015}}           &  $\checkmark$ &  $\checkmark$ &       $\times$             &   $\times$&     High  &  Low &   \begin{tabular}[c]{@{}c@{}} PC \\  \end{tabular}       & \begin{tabular}[c]{@{}c@{}}Depends \\ on App\end{tabular}  & N/A    \\ \hline
\multicolumn{1}{|c|}{\textbf{Sora}\cite{Tan2011}}          &    $\checkmark$                      &          $\checkmark$            &       $\times$              &      $\times$              &            High              &        Low                    &      PC         &  \begin{tabular}[c]{@{}c@{}}Xilinx \\ Virtex-5\end{tabular}  & \begin{tabular}[c]{@{}c@{}}$\sim900$ \\ Board Only\end{tabular}   \\ \hline
\multicolumn{1}{|c|}{\textbf{SODA}\cite{Lin2006}}          &  $\checkmark$&  $\checkmark$ & $\checkmark$ &  $\times$&  High&  High& \begin{tabular}[c]{@{}c@{}}ARM Cortex-M3 \\ + Processing Elements\end{tabular}& N/A & N/A  \\ \hline
\multicolumn{1}{|c|}{\textbf{Iris}\cite{Sutton2010}}          & $\checkmark$  &  $\checkmark$ &  $\checkmark$&  $\checkmark$& High&   High& \begin{tabular}[c]{@{}c@{}}Dual-core ARM \\ Cortex-A9\end{tabular} &  \begin{tabular}[c]{@{}c@{}}Xilinx \\ Kintex-4\end{tabular} &    \begin{tabular}[c]{@{}c@{}}$\sim1.2K$ \\ Total\end{tabular} \\ \hline
\multicolumn{1}{|c|}{\textbf{Atomix}\cite{Bansal2015}}        &  $\checkmark$   & $\checkmark$ &  $\checkmark$&  $\checkmark$&  High&    Medium        &   TI 6670 DSP&  N/A   &  \begin{tabular}[c]{@{}c@{}}$\sim200$ \\ DSP Only\end{tabular} \\ \hline
\multicolumn{1}{|c|}{\textbf{BeagleBoard-X15}\cite{beagle}}        &  $\checkmark$   & $\checkmark$ &  $\checkmark$&  $\checkmark$&  High&    Medium        & \begin{tabular}[c]{@{}c@{}}$2\times$ TI C66x DSPs +\\ $2\times$ ARM Cortex-A15 \& $2\times$ M4\end{tabular}&  N/A   &  \begin{tabular}[c]{@{}c@{}}$\sim270$ \\ Board Only\end{tabular} \\ \hline
\multicolumn{1}{|c|}{\textbf{Airblue}\cite{Ng2010}}       & $\checkmark$&                     $\checkmark$ &   $\checkmark$ &  $\checkmark$ &  High &  High &  N/A   & \begin{tabular}[c]{@{}c@{}}Intel \\ Cyclone IV\end{tabular}  &  \begin{tabular}[c]{@{}c@{}}$\sim1.3K$ \\ Board Only\end{tabular} \\ \hline
\multicolumn{1}{|c|}{\textbf{WARP v3}\cite{warp}}           &   $\checkmark$                       &       $\times$        &      $\checkmark$    &         $\checkmark$                                  &     High   &  High &    \begin{tabular}[c]{@{}c@{}}$2\times$ Xilinx \\ MicroBlaze cores\end{tabular}             & \begin{tabular}[c]{@{}c@{}}Xilinx \\ Virtex-6\end{tabular}                                          & \begin{tabular}[c]{@{}c@{}}$\sim7K$ \\ Total\end{tabular}   \\ \hline
%
%
%
\multicolumn{1}{|c|}{\textbf{PSoC 5LP}\cite{psoc}}           &                         $\checkmark$  &      $\times$               &    $\checkmark$                  &                    $\checkmark$ &         Low                 &             High               &      ARM Cortex-M3           & N/A & \begin{tabular}[c]{@{}c@{}}10 \\ Board Only\end{tabular}  \\ \hline
\multicolumn{1}{|c|}{\textbf{Zynq-based}\cite{Drozdenko2017}}           &                         $\checkmark$  &      $\checkmark$                &     $\checkmark$                 &                    $\checkmark$ &           High               &         High                   &     \begin{tabular}[c]{@{}c@{}}Dual-core ARM \\ Cortex-A9\end{tabular}   & \begin{tabular}[c]{@{}c@{}}Xilinx \\ Kintex-4\end{tabular} &  \begin{tabular}[c]{@{}c@{}}$\sim1.2K$ \\ Total\end{tabular} \\ \hline
\end{tabular}
\end{table*}

\section{Related Research and Potential Solutions}
\label{related}

Although in the previous sections we have highlighted some of the challenges of building SDRs, in this section, we present additional research areas that are still being faced by the research and development community. 
These challenges are technical or practical.

\subsection{Remote System Update}

One of the main features and motivations of SDRs is their reconfigurability and flexibility. 
In order to take full advantage of this, the process of updating a SDR platform should be quick and easy.  
Remote stand alone SDRs are usually FPGA and DSP based, with more FPGAs being used. 
Hence, most of the research has been focused on remote updates of FPGAs. 
Since FPGAs are volatile, which are configured during system power-on through their flash memories, update is traditionally done using Joint Test Action Group (JTAG)  method \cite{ZhangXin2015}. 
However, with more SDRs deployed and adopted in wireless and cellular networks, remote update becomes necessary and a challenge. 
With remote over the air (OTA) updates, it becomes possible to send patches to current designs, or even upload an entirely new and improved design to mobile networks and BSs \cite{Hoffmeyer2004,Bing2006}. 
Some of the challenges faced by the research community include speed, reliability, cost, and security \cite{Zhou2017}. 

In \cite{ZhangXin2015}, the authors introduce a method based on RS-422 serial communication and High Level Data Link Control (HDLC) protocol to update DSP and FPGA systems. 
Their method is fast, stable and easy to implement. 
The authors in \cite{Fernandes2016} show a new method for remotely updating Xilinx FPGAs \cite{xilinx} by storing the new design or code in Serial Peripheral Interface (SPI) flash memories. 
They utilized the Xilinx Quick Boot application, along with the KC705 Evaluation Kit from Xilinx, in order to develop their method. 
However, it is not always practical or feasible to use a download cable in order to update the system. 
The authors in \cite{Zhou2017} tackle this issue and the problem of downtime during an update or in the case of a failure. 
With this method, there exists two images, namely a factory mode configuration image, which acts as the baseline, and an application mode configuration image, which is used for some specific functions. 
They show the capability of updating and running a new application mode (design) image, in addition to rolling back into the factory mode image when no application mode image is available. 
Others have focused their efforts on improving the security of remote updates, such as \cite{Vliegen2014} and \cite{Kashyap2016}. While there exists some solutions to the remote update process, a few challenges including partial reconfiguration of FPGAs are not yet resolved.

\subsection{Centralized Algorithms and Network Slicing}
In order to simplify the design and provisioning of large-scale networks, software-defined networking (SDN) has been proposed to centralized network control.
In this architecture, a controller (or multiple controllers) communicate with network devices (data plane) to collect their status information and configure their operation.
Recent studies show that wireless networks can significantly benefit from the central network view established in the controller to design more efficient network control algorithms such as channel assignment and mobility control.

Centralized control of network resources is the enabler of network slicing, which refers to the abstraction, slicing and sharing of network resources.
Three levels of slicing is applicable to wireless networks: 
(i) spectrum (a.k.a., link virtualization): requires frequency, time or space multiplexing. 
(ii) infrastructure: the slicing of physical devices such as BSs. 
(iii) network: this refers to the slicing of the network infrastructure.
Compared to wired networks, slicing the resources of wireless networks is significantly more challenging due to the variable nature of wireless channel.
Meanwhile, since SDRs enable the slicing of resources both at the spectrum and infrastructure level, they can be used to augment SDNs towards a fine-grain allocation of resources.
For example, compared to ASIC-based transceivers, SDRs provide a significantly higher level of control over the parameters of physical and MAC layer.

It should be noted that when centralized network control is employed, the delay of communication between the controller and SDR platforms as well as the delay of programming and applying new configurations should be bounded within the specification requirements.
For example, in a dense and mobile environment, the controller may employ a centralized channel and power control algorithm to instruct the nodes adjust their channels based on the decisions made centrally.
In this case, it is essential to ensure the delay of sending configuration messages to multiple SDR platforms meets the requirements of central algorithm. 
Furthermore, it is essential to ensure all the SDRs apply the configuration at the same time, otherwise serious interference and collision might happen.
Although protocols such as OpenFlow \cite{OpenFlow} and Netconf \cite{enns2011network} have been designed for interactions between the controller and data plane, their implications on the performance of wireless networks have not been investigated.
Specifically, it is essential to evaluate the effect of hardware platform (i.e., GPP, DSP, FPGA) on the delay of applying configurations.
These challenges have not been addressed yet.

\subsection{Network Functions Virtualization (NFV)}

One of the new topics is the concept of Network Functions Virtualization (NFV), which offers an alternative way of designing and managing networking functions. 
The concept is very similar to SDRs, in the sense that various network functions can run in software on top of different hardware platforms. 
These platforms are typically high-volume servers, storage devices, and cloud-computing data centers \cite{Mijumbi2016}. 
Some of the functions that are virtualized via this method are load balancing, firewalls, intrusion detection devices and WAN accelerators. 
This flexibility is what makes NFV very attractive for network operators, carriers, and manufacturers, in addition to cost reductions \cite{GilHerrera2016}. 
From the SDR point of view, instead of performing signal processing operations on the platform, these operations are offloaded to a general computing platform. 
Such an architecture reduces the load of edge devices and BSs as can benefit from powerful processors to implement complex signal processing operations.
For example, when multiple SDRs are connected to a computing platform, sophisticated signal processing algorithms could be developed to cope with challenges such as interference.


To leverage NFV for SDRs, developers have been attempting to tie them together to achieve complete flexibility across the platform \cite{Sun2015,Shome2015}. 
Although several wireless SDN architectures have been proposed to address the challenges of wireless communication \cite{OpenSDWN,BIGAP,Primitives,CloudMAC3}, most of them do not benefit from the features of SDRs.

\subsection{Energy Efficiency}

Battery-powered devices in an IoT network face the challenge of minimizing power consumption in order to extend battery-life before they are due for a replacement, which is a costly operation. 
Ready-to-deploy SDR systems may use high-performing platforms, such as FPGAs, without providing solutions or alternatives to batteries \cite{Ang2013}. 
Even in the case of BSs with on-grid power sources, it has become crucial to lower power consumption in order to reduce $CO_{2}$ emissions \cite{Han2016}. 
This is particularly important for cellular network operators, where BSs consume more that 50\% of the total energy consumed in the network \cite{Han2013}. 
To address these concerns, energy harvesting mechanisms have been introduced.
Energy harvesting or scavenging is the process of deriving power from external sources, such as solar and wind energies (also, referred to as green energy), and stored for consumption alongside internal sources (e.g., battery or electrical grid sources) \cite{Tentzeris2014}. 
As this green energy is a viable option for powering BSs \cite{Han2014}, Ericsson (a major telecommunication company) has invested into and designed green-energy-powered BSs motivated by environmental and financial reasons \cite{EricssonInc.2017}. 
Therefore, hybrid power operation has been accepted as a solution to lower electrical grid energy consumption and cost \cite{Han2013}.

In \cite{Tentzeris2014}, the authors present a survey on energy harvesting technologies with regards to transducers, such as antennas and solar cells, that can utilize renewable energy sources, and cover some of the applications in the IoT and M2M world. 
The authors in \cite{Han2014} overview the cellular network operators that have adopted the hybrid solution and started using green energies to power their BSs. 
The authors in \cite{Park2013} discuss the issue of implementing an energy harvesting transmitter in a cognitive radio.
They also derive the optimal spectrum sensing policy that maximizes the expected total throughput under energy causality and collision constraints. 
Energy causality states that the total consumed energy should be equal to or less than the total harvested energy, whereas the collision constraint states that the probability of accessing the occupied spectrum is equal to or less than the target
probability of a collision.
The authors concluded that a battery-operated radio can operate for a long time by optimizing both the energy and spectral efficiency.
Energy harvesting is often associated with what is known as "Green IoT", which is the new trend for IoT networks and devices, where the main focus is making them more energy efficient. 
Another relevant work is \cite{Shaikh2017,Arshad2017}, where the authors present an overview of the challenges and existing solutions of energy-efficient IoT systems.

\subsection{Co-Design}

By definition, co-design is the process of realizing system-level goals through exploiting the trade-offs between software and hardware in an integrated design. 
This yields a higher design quality, and optimizes the cost and design cycle time, which in turn shortens the time-to-market time. 
As co-design is adopted for more applications, developers typically face the problem of partitioning and scheduling. 
Finding the optimal design partition is not trivial.
While system profilers can assist in providing an insight about the system's behavior and help identify the parts of the code that can be accelerated on hardware, partitioning should be an automatic process and requires no external involvement.
There are several algorithms proposed to address the challenges of partitioning, such as PSO, FCM, and FCMPSO \cite{Mhadhbi2016}. 
These are optimization algorithms used to mapping embedded applications to Directed Acyclic Graphs (DAGs) for multi-core architectures. However, in SDRs, the problem is even more complicated due to having two layers of operation, namely PHY and MAC, and partitioning needs to take into account strict real-time requirements.
It is a delicate equilibrium between performance, cost, and correct operation.
Even with the process being more challenging than other design approaches, the benefits of co-design for complex systems outweigh the initial cost.

\subsection{Security}
\label{security}
SDRs simplify security provisioning.
For example, when a new security mechanism does not require hardware replacement (e.g., 802.11i's WPA), it can be implemented through reprogramming SDRs.
On the other side, the reprogrammability feature of SDRs exposes security threats, whether they are standalone or part of a SDN architecture.
For example, assume an 802.11 network employs SDR-based BSs (i.e., access points).
In this case, a denial of service (DoS) attack can be implemented by instructing a large number of clients to associate with a BS.
Also, if the controller is compromised, then all the SDRs might be reprogrammed to be nonfunctional.
Therefore, it is important to identify security threats and take them into account when designing SDR-based wireless networks.

Offloading SDR processing to general processing platforms through NFV enables the deployment of sophisticated central algorithms for detecting abnormal activities and network breaches.
For example, by analyzing the signal strength received from a client at one or multiple BSs, a denial of service attack caused by generating excessive interference could be detected.
Proposing security mechanisms that benefit from the features of SDRs is an important future direction, especially for large-scale and public networks.

\section{Existing Surveys}
\label{old}

Joseph Mitola III was the first to use the term "Software Radio" in 1993, and published an important work introducing and explaining the concept of using software rather than traditionally used hardware for designing radio systems \cite{Mitola93}. 
In an interesting early survey (1999) \cite{Tuttlebee1999} on the "then" emerging technology, namely SDR, the authors made the case that SDRs have a great potential to  advance and facilitate the development of communication standards such as WiFi and cellular networks. 
In a review published a few years later \cite{Haghighat2002a}, the author paid close attention to the hardware component of SDRs, such as programmable RF modules and high-performance DACs and ADCs, as more technological advancement had been made.

There are very few notable works to survey and review different SDR platforms and testbeds. 
One such survey is the work by \cite{Ulversoy2010}, where the author discusses the challenges SDR developers face. 
These challenges include size, weight, power, software architecture, security, certification and business opportunities. 
While this work is important for compiling information about these challenges and presenting them in one place, it stays away from enumerating the different SDR platforms, implementation approaches and their applications in the communication standards world.

In that regard, the authors of \cite{Dardaillon2012} compare the SDR platforms developed by the year 2012 and give a brief description of what each platform entails. 
It lacks, however, any detailed comparison based on the different categories of computational power, energy efficiency, flexibility, adaptability, and cost. 
It is through these comparisons that a designer is able to make an informed decision on what platform to adopt for their specific application. 
The authors of \cite{Anjum2011} attempt to list and review several DSP-based SDR platforms from both academia and industry, with more focus on commercial solutions, and then providing a simple comparison between them in terms of programmability, power, and flexibility. 
However, what this work lacks, in addition to being outdated, is the discussion of FPGAs (mentioned only one example) and hardware/software co-design solutions, and a methodical analysis of different design approaches based on a set of performance metrics.

Authors of \cite{Cummings2007} presented a survey of a few SDR platforms that had been developed more than a decade ago. 
The authors of \cite{moy2015} presented a paper that laid out the development and evolution of SDRs over the past several years and discussed the motivation behind gaining even more attention recently. 
Another work is the comparison conducted by the authors in \cite{Palkovic2010} between the Imec Bear platform \cite{imec} and a few multicore-based SDR platforms. 
Other attempts include the work by \cite{Machado2015}, where the authors focused on discussing the analog end of the SDR concerning signal sampling, processing, and the hardware/software responsible for handling these tasks. 
The work by \cite{Wyglinski2016} compares several SDR platforms to prove the feasibility and reliability of using SDRs in education, industry, and government. 
Another outdated survey is \cite{Farrell2009b}.

Other surveys that are relevant to SDR platforms is the work by \cite{Nesimoglu2010a}, where the author discusses and reviews the state-of-the-art in microwave technology in transceivers. 
The paper explores the development of SDRs using different technologies in radio frequency engineering. 
It is a comprehensive study of several research topics such as the design of tunable radio frequency components, linear and power efficient amplifiers, linear mixers, and interference rejection techniques. 
Whereas, authors in \cite{Baldini2012} provide a very thorough study on the several security threats and challenges in SDRs, and go over their certification process. 
As SDR platforms grow in popularity, and more communication protocols are realized using them, it becomes essential that developers are prepared for security issues and well-equipped with tools that protect their systems.

\section{Conclusion}
\label{conc}

In this paper, we provided a comprehensive overview of the various design approaches and hardware platforms adopted for SDR solutions. 
This includes GPPs, GPUs, DSPs, FPGAs, and co-design. 
We explained the basic architectures and analyzed their advantages and disadvantages. 
Due to the different features of design approaches, it was important to compare them against each other in terms of computational power and power efficiency. 
We then reviewed the major current and past SDR platforms, whether they were developed by the industry or in academia. 
Finally, we discussed some of the research challenges and topics that are predicted to improve in the near future, helping to advance SDRs and their wide adoption.

We believe that SDR solutions are going to be mainstream and that their ability to implement different wireless communication standards with high levels of flexibility and re-programmability will be considered the norm. 
This paper poses as an exhaustive overview of this phenomenon, its enabling technologies, applications, and the current research tackling this issue from different angles. 





\ifCLASSOPTIONcaptionsoff
  \newpage
\fi

\Urlmuskip=0mu plus 1mu\relax
\bibliographystyle{IEEEtran}
\bibliography{library} 
\end{document}